\documentclass[pageno]{jpaper}

\newcommand{\ignore}[1]{}

\usepackage[normalem]{ulem}
\usepackage{amsmath,amsfonts}

\usepackage[normalem]{ulem}
\usepackage{listings}
\usepackage{amsmath}
\usepackage[noend,linesnumbered,ruled,vlined]{algorithm2e}
\usepackage{physics}
\usepackage{amsmath}
\usepackage{multirow}
\usepackage{enumitem}
\usepackage{amssymb}
\usepackage{adjustbox}
\usepackage[normalem]{ulem}
\usepackage[flushleft]{threeparttable}
\usepackage{dblfloatfix}

\usepackage{graphicx}
\usepackage{subfig}

\SetCommentSty{mycommfont}

\def\BibTeX{{\rm B\kern-.05em{\sc i\kern-.025em b}\kern-.08em
    T\kern-.1667em\lower.7ex\hbox{E}\kern-.125emX}}

\usepackage{url}

\usepackage[noadjust]{cite}
\usepackage{tikz}

\def\BibTeX{{\rm B\kern-.05em{\sc i\kern-.025em b}\kern-.08em
    T\kern-.1667em\lower.7ex\hbox{E}\kern-.125emX}}
\definecolor{aliceblue}{rgb}{0.94, 0.97, 1.0}
\usepackage{xcolor}
\usepackage{tcolorbox}
\tcbset{width=1\columnwidth,boxrule=1pt,colback=aliceblue,arc=1pt,left=2pt,right=2pt,boxsep=0.5pt}

\usepackage{authblk}

\begin{document}

\author[1,*]{Ramin Ayanzadeh}
\author[1]{Poulami Das}
\author[2]{Swamit S. Tannu}
\author[1]{Moinuddin Qureshi}

\affil[1]{Georgia Institute of Technology, Atlanta, GA, USA}
\affil[2]{University of Wisconsin--Madison, Madison, WI, USA}

\affil[*]{ayanzadeh@gatech.edu}

\title{
{EQUAL}: Improving the Fidelity of Quantum Annealers\\ by Injecting Controlled Perturbations
}

\date{}
\maketitle

\thispagestyle{empty}

\begin{abstract}

   Quantum computing is an information processing paradigm that uses quantum-mechanical properties to speedup computationally hard problems. Gate-based quantum computers and Quantum Annealers (QAs) are two commercially available hardware platforms that are accessible to users today. Although promising, existing gate-based quantum computers consist of only a few dozen qubits and are not large enough for most applications. On the other hand, existing QAs with few thousand of qubits have the potential to solve some domain-specific optimization problems. QAs are single instruction machines and to execute a program, the problem is cast to a Hamiltonian, embedded on the hardware, and a single quantum machine instruction (QMI) is run. Unfortunately, noise and imperfections in hardware result in sub-optimal solutions on QAs even if the QMI is run for thousands of trials. 
       
   The limited programmability of QAs mean that the user executes the same QMI for all trials. This subjects all trials to a similar noise profile throughout the execution, resulting in a \emph{systematic bias}. We observe that systematic bias leads to sub-optimal solutions and cannot be alleviated by executing more trials or using existing error-mitigation schemes. To address this challenge, we propose \emph{EQUAL} (\underline{E}nsemble \underline{QU}antum \underline{A}nnea\underline{L}ing). EQUAL generates an ensemble of QMIs by adding controlled perturbations to the program QMI. When executed on the QA, the ensemble of QMIs steers the program away from encountering the same bias during all trials and thus, improves the quality of solutions. Our evaluations using the 2041-qubit D-Wave QA show that EQUAL bridges the  difference between the baseline and the ideal by an average of 14\% (and up to 26\%), without requiring any additional trials. EQUAL can be combined with existing error mitigation schemes to further bridge the difference between the baseline and ideal by an average of 55\% (and up to 68\%).
       
\end{abstract}

\section{Introduction}

Quantum computing is an information processing paradigm that leverages quantum mechanical properties of quantum bits (qubits) to store and process information and promises significant computational advantages for many hard problems~\cite{shor1999polynomial,grover1996fast,feynman1982simulating,lloyd1996universal}.
There exist different models for the physical realization of this computational paradigm such as gate model quantum computers and quantum annealers \cite{nielsen2010quantum,albash2018adiabatic}. 
Currently, prototypes of both gate model~\cite{IBMQ,GoogleAI,honeywell} and annealing~\cite{D-Wave} types are available and some of them can already outperform modern supercomputers for some tasks~\cite{arute2019quantum,villalonga2020establishing,king2021scaling,wu2021strong}. 

\begin{figure*}[ht]	
	\centering
	\includegraphics[width=\linewidth]{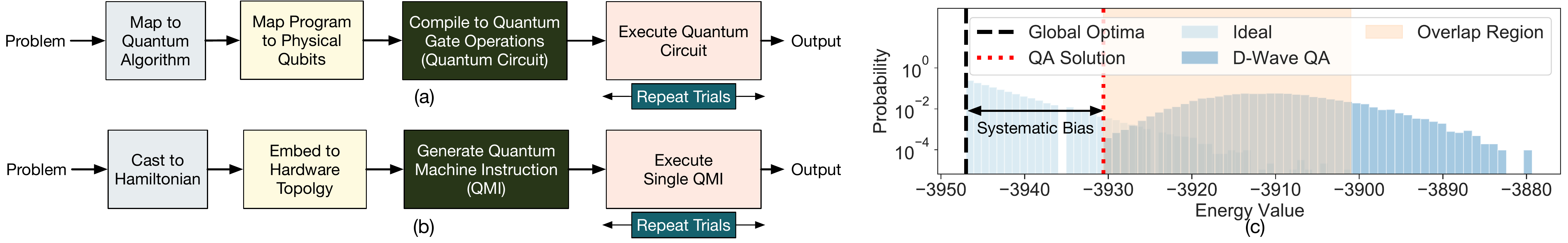}
	\vspace{0.1in}
    \caption{Steps involved in solving a problem using (a) gate-model quantum computers and (b) quantum annealers. 
	(c) Energy histogram of a 2000-qubit optimization benchmark executed on D-Wave QA (in logscale). The QA can quickly identify the region of the ground state energy (overlapping region) but the solution is far from the global optima due to systematic bias.
	}\vspace{0.1in}
	\label{fig:intro}
\end{figure*}

Gate-based quantum computers, such as IBM and Google machines, use discrete quantum gate operations to manipulate qubits such that the state of the qubits evolve to produce the desired outcome as the program proceeds. Such systems with about 50-plus qubits are already available~\cite{AmazonBraKet,IBMQ,MicrosoftAzure}. 
To solve a problem on a gate-based quantum computer, we map it to an efficient quantum algorithm, map the high-level program qubits to the physical qubits of the device, translate the instructions into a series of low-level quantum gates, and execute them, as shown in Figure~\ref{fig:intro}(a). 
Although these types of systems promise significant computational advantages in the near-term, they must grow in size for practical applications~\cite{nielsen2010quantum,villalonga2020establishing,preskillNISQ}. 

Unlike gate model quantum computers that can be programmed to solve different classes of problems, \emph{Quantum Annealers (QAs)} are single-instruction machines that can only solve a specific discrete optimization problem by 
minimizing the energy of a physical system, called \emph{Hamiltonian}~\cite{das2008colloquium,albash2018adiabatic}.  
To solve a problem on a QA, (a)~we cast it to a Hamiltonian, (b)~embed it to match the topology of the QA device, (c)~obtain the resulting single \emph{Quantum Machine Instruction (QMI)}, (d)~execute the single QMI, and (e) repeatedly run the same QMI multiple times~\cite{mcgeoch2020theory}, as shown in Figure~\ref{fig:intro}(b). 
The outcome with the lowest energy is deemed as the solution.

As QAs can only minimize a specific objective function, any other problem must be cast/reduced to this Hamiltonian. 
\emph{Casting} computes the coefficients of the Hamiltonian such that the global minima of the Hamiltonian represents the global optima of the problem of interest~\cite{lucas2014ising,ayanzadeh2020reinforcement,mcgeoch2020theory}. 
\emph{Embedding} maps the problem graph to the topology of the QA. 
As QAs have limited connectivity between qubits, embedding encodes a program qubit with higher connectivity by using a chain of physical qubits. 
The problem of limited connectivity exists even on most existing gate-based quantum computers and can be overcome by inserting SWAP operations~\cite{zulehner2018efficient,murali2019noise,tannu2019not}. 
However, a similar approach is impractical for QAs as they can only execute a single QMI. Unlike gate-based systems, QAs available today with 5000-plus qubits~\cite{D-Wave,AmazonBraKet,mcgeochd} are much larger, scale faster,
and have the potential to power a wide range of real-world applications---including, but not limited to,
planning~\cite{rieffel2015case}, scheduling~\cite{venturelli2015quantum,tran2016hybrid}, constraint satisfaction problems~\cite{bian2016mapping}, Boolean satisfiability (SAT)~\cite{su2016quantum,ayanzadeh2020reinforcement}, matrix factorization~\cite{o2018nonnegative}, cryptography~\cite{peng2019factoring,hu2020quantum}, fault detection and system diagnosis~\cite{perdomo2015quantum}, compressive sensing~\cite{ayanzadeh2019quantum,ayanzadeh2020ensemble}, control of automated vehicles~\cite{inoue2021traffic}, finance~\cite{elsokkary2017financial}, material design~\cite{kitai2020designing}, and protein folding~\cite{mulligan2020designing}. 
Although promising, QA hardware suffers from various drawbacks such as noise, device errors, limited programmability and low annealing time, which degrade their reliability~\cite{albash2018adiabatic,mcgeoch2020theory,ayanzadeh2020multi}. 
Addressing these limitations requires device-level enhancements that may span generations of QAs. 
Therefore,  software techniques to improve the reliability of QAs is an important area of research~\cite{cai2014practical,date2019efficiently,golden2019pre,borle2019post,goodrich2018optimizing,okada2019improving,D-Wave_Ocean_SDK}.

Despite recent hardware and software enhancements, existing QAs may fail to find the global minima for certain  problems~\cite{ayanzadeh2020multi}. 
For example, Figure~\ref{fig:intro}(c) shows the energy histogram of a 2000-qubit optimization problem on a D-Wave QA. 
We can think of QA as a machine that samples from a Boltzmann distribution such that samples with lower energy values are exponentially more likely to be observed~\cite{vinci2016nested,ayanzadeh2020multi}. 
In theory, QA can find the optimal solution with a very high probability~\cite{nishimori2017exponential}. 
However, in this example, we observe that although the QA can quickly identify the region of the global optima, the best solution from the QA is far from the global optima. 
As users run only a single QMI, the program is subjected to a similar noise profile for all trials, resulting in a \emph{systematic bias}. 
Our experiments show that running more trials or relying on existing error-mitigation schemes cannot overcome this bias. 
Unfortunately, systematic bias produce incorrect solutions far from the global optima and limits the reliability of QAs for practical applications.

In this paper, we propose \emph{Ensemble Quantum Annealing (EQUAL)}, an effective scheme for mitigating systematic bias and improving the reliability of QAs by running an ensemble of QMIs with controlled perturbations. 
EQUAL is based on the insight that running the same QMI for all trials projects QAs to a very similar noise profile and bias. 
Instead, EQUAL uses an ensemble of QMIs that subjects the system to different noise profiles and biases. 
Generating effective ensembles of QMIs is non-trivial and our design focuses on addressing it. 
\footnote{The problem of systematic bias is similar to correlated errors on gate-model quantum computers that can be addressed by mapping programs to different sets of physical qubits on the same machine~\cite{tannu2019ensemble}, inserting different SWAP routes ~\cite{tannu2019ensemble}, or using different machines~\cite{patel2020veritas}. The equivalent step for QAs would be to use multiple embeddings. However, this is not viable for QA and we discuss the details in Section~\ref{sec:related_work}.}

To generate ensembles, EQUAL creates new Hamiltonians, called \emph{Perturbation Hamiltonians}, 
and adds them to the original problem Hamiltonian. 
Every perturbation Hamiltonian adds noise to the original Hamiltonian and the QMI obtained from this process is a perturbed variation of the original QMI.  
The challenge in this step is that adding extremely small perturbations will have no impact on the systematic bias, whereas adding large perturbations can significantly change the landscape of the original Hamiltonian. 
In the worst case,  the final perturbed Hamiltonian may correspond to a problem completely different from the one at hand. 
Thus, there exists a trade-off between the ability to eliminate systematic bias and the correctness of a Hamiltonian. 
Ideally, we want a perturbed Hamiltonian that can eliminate systematic bias without altering the characteristics of the problem Hamiltonian significantly. 
To address this challenge, EQUAL exploits the fact that QAs only allow a limited precision of coefficients for a Hamiltonian due to hardware limitations. 
For every ensemble, EQUAL draws the coefficients of the corresponding perturbation Hamiltonian randomly at a range just below the supported precision so that adding the Perturbation Hamiltonian may only shift the coefficients of the QMI (post truncation) to one of the neighboring quantization levels and not impose significant changes to the landscape of the original Hamiltonian.

We also analyze existing error-mitigation approaches for QAs. 
Our characterization experiments on D-Wave shows that the SQC~\cite{ayanzadeh2020multi} post-processing technique is highly effective for D-Wave.  
We compare EQUAL with SQC and show that the two schemes can be combined for even greater benefit.  
The resulting design, \emph{EQUAL+}, provides significantly better fidelity than EQUAL and SQC standalone. 
As the SQC post-processing relies only on classical computations, EQUAL+ does not incur any additional trials compared to EQUAL.

Our evaluations on D-Wave's 2041-qubit QA show that EQUAL bridges the  difference between the baseline and the ideal by an average of 14\% (and up to 26\%). 
EQUAL+ further bridges the difference between the baseline and the ideal by an average of 55\% (and up to 68\%).

\vspace{0.05 in}
Overall, this paper makes the following contributions:
\vspace{0.05 in}
\begin{enumerate}[leftmargin=0cm,itemindent=.5cm,labelwidth=\itemindent,labelsep=0cm,align=left, itemsep=0.15cm, listparindent=0.3cm]
	\item We show that there is a systematic bias associated with each QMI, running on QA, that deviates the annealing process from achieving the ground state of the corresponding Hamiltonian and produces sub-optimal solutions.
	\item We propose {\em EQUAL (Ensemble Quantum Annealing)} to mitigate the bias by forming multiple perturbed copies of a given QMI and running each for a subset of trials. 
	
	\item We propose an effective method to generate the perturbed copies while retaining the structure of the problem by leveraging the hardware imperfections from limited precision.
	
	\item We propose EQUAL+ that combines EQUAL with existing SQC error-mitigation to further improve the reliability.
	
\end{enumerate}

\section{Background and Motivation}

\subsection{Quantum Computing}
Quantum computing is a computational paradigm that stores and processes information using quantum bits or \emph{qubits}. 
The state of a qubit $\ket{\psi}$ can be represented as a superposition of its two basis states $\ket{0}$ and $\ket{1}$ using a vector: $\ket{\psi} = \alpha\ket{0} + \beta\ket{1}$, 
where $\alpha$ and $\beta$ are complex probability amplitudes associated with the basis states. 
Similarly, an $N$-qubit system exists in a superposition of $2^{N}$ basis states. 
This exponential scaling in state space with a linear increase in qubits enables quantum advantage. 
Currently, two types of quantum platforms are available to users through cloud services~\cite{AmazonBraKet,IBMQ,MicrosoftAzure}---gate-based quantum computers and quantum annealers.  

\vspace{0.05in}
{\noindent \textbf{Gate-based Quantum Computers:}} 
A gate-based quantum computer executes a predefined sequence of quantum gate operations to transform the initial state of the qubits to the desired state by changing the superposition. 
The sequence of quantum gates is known as a quantum circuit. 
Quantum computers from IBM, Google, and others use gate-based model.

\begin{figure}[tp]
\centering
\includegraphics[width=\columnwidth]{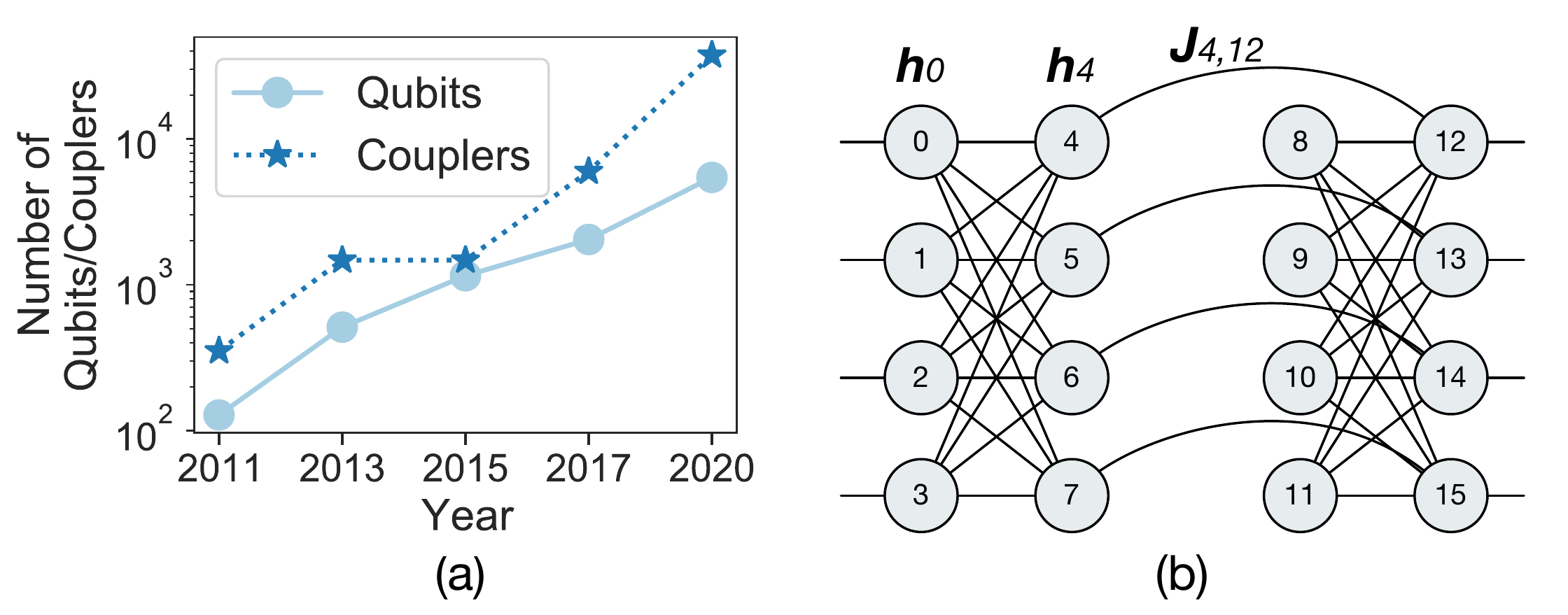}
\caption{
	(a) Evolution of Quantum Annealers (QAs) over time 
	(b) A cropped view of the D-Wave 2041-qubit QA with its connectivity graph where the nodes denote qubits and edges represent couplers (or connectivity between two qubits). 
}
\label{fig:DW_Chimera}
\end{figure}

\vspace{0.05in}
{\noindent \textbf{Quantum Annealers:}} 
Quantum annealing is a meta-heuristic for solving combinatorial optimization problems that runs on classical computers~\cite{amara1993global,finnila1994quantum,kadowaki1998quantum,das2008colloquium,ohzeki2011quantum,nishimori2017exponential}. 
Quantum Annealers are single instruction machines for solving combinatorial optimization problems.  
Unlike gate model quantum computers, where we directly change the state of qubits via quantum gates, QAs control the environment, and qubits evolve to remain in the ground state (i.e., a configuration with the lowest energy value) of a Hamiltonian (or energy/cost function)~\cite{das2008colloquium,albash2018adiabatic}. 
Quantum annealers, such as the ones from D-Wave, are analog systems that can only minimize the following energy function:
\begin{equation}
	\label{eqn:DW_E}
	\mathbf{H}_p := f(\mathbf{z}) =  \sum_{i=0}^{N-1}{\mathbf{h}_i \mathbf{z}_i} + \sum_{i=0}^{N-1}{\sum_{j=i+1}^{N-1}{J_{ij} \mathbf{z}_i \mathbf{z}_j}},
\end{equation}
where $N$ is the number of qubits, 
$\mathbf{h}_i \in \mathbb{R}$ specifies the linear coefficient of qubit ${i},$ 
$J_{i,j} \in \mathbb{R}$ represents the coupler weight between qubits ${i}$ and ${j},$ 
and $\mathbf{z}_i$ is the variable that can take its value from $\{-1,+1\}$~\cite{mcgeoch2020theory,ayanzadeh2020multi,ayanzadeh2020reinforcement}. 
Ever since the introduction in 2011, QAs have rapidly scaled in size up to few thousand of qubits, as shown in Figure~\ref{fig:DW_Chimera}(a) and promise significant computational advantage for a wide range of applications.

\subsection{Operation Model of QA}
To execute a program on a QA, the problem is cast to a Hamiltonian such that its global minimum represents the optimal solution of the problem at hand. 
This step computes the coefficients $\mathbf{h}$ and ${J}$, denoted in Equation~\eqref{eqn:DW_E}, corresponding to the \emph{quantum machine instruction (QMI)} to be executed on the QA. 
Executing the QMI on a QA returns a sample $\mathbf{z} = \{\mathbf{z}_0, \mathbf{z}_1, \dots, \mathbf{z}_{N-1} \}$ 
as a potential minimum of the corresponding energy function. 
Unfortunately, executing a QMI only once may not result in the ground state of the Hamiltonian due to noise in the system~\cite{ayanzadeh2020multi}. 
Thus, in practice, the process of executing the single QMI is repeated for thousands of trials.  The sample with the lowest energy is reported as the solution.

\subsection{The Opportunity: 
Solving Large Problems with QA} 
\begin{figure}[ht!]
	\centering
	\includegraphics[width=\columnwidth]{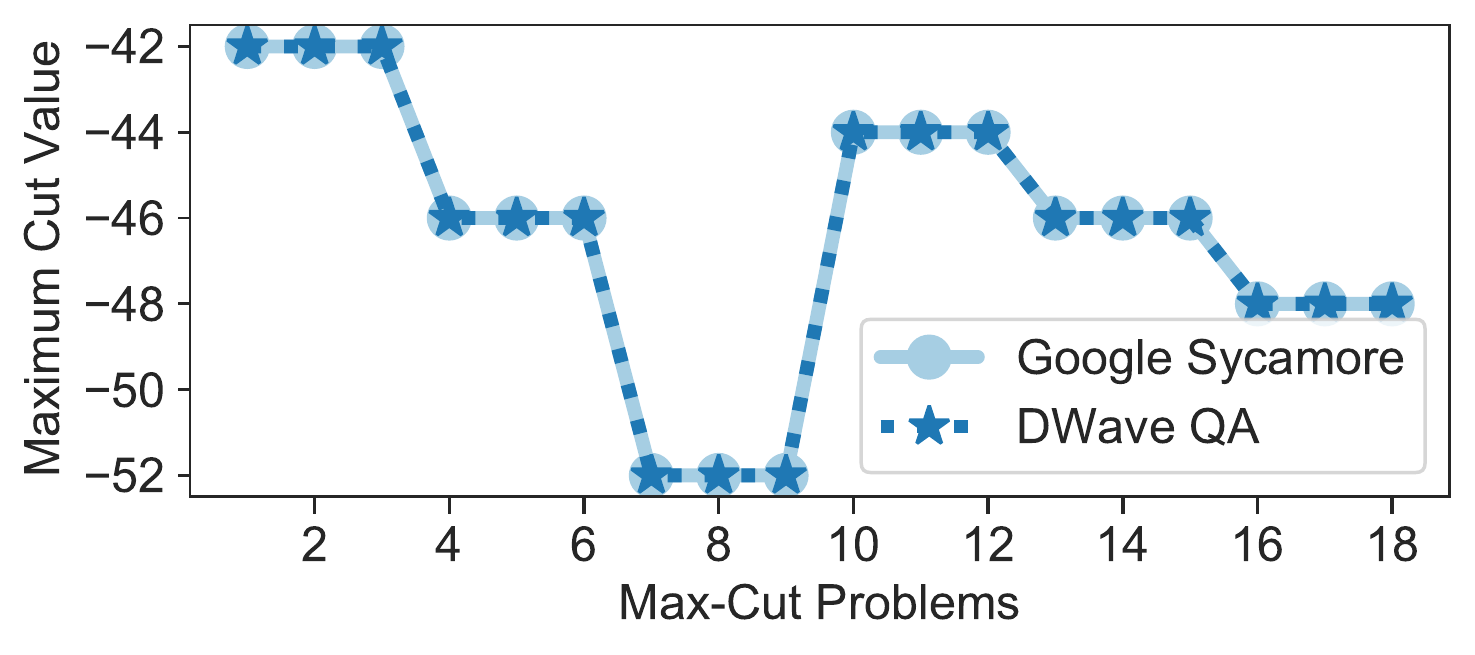}
	
	\caption{Comparison of 54-qubit Google Sycamore~\cite{harrigan2021quantum} and 2041-qubit D-Wave Quantum Annealer.}
	\label{fig:googledwavecompare}
\end{figure}

Google Sycamore is a state-of-the-art 54-qubit gate-based quantum computer that can outperform even the most powerful supercomputer for some tasks~\cite{arute2019quantum}. 
We compare the performance of the D-Wave 2041-qubit QA and Google Sycamore for 18 different Max-Cut problems. 
The Max-cut problems used in this evaluation correspond to the fully-connected Sherrington-Kirkpatrick (SK) Model~\cite{sherrington1975solvable} and uses up to 17 qubits~\cite{harrigan2021quantum,quantum_ai_team_and_collaborators_2020_4091470,Google_AI_Quantum_and_Collaborators2020}. 
These are some of the hardest benchmarks on Sycamore as fully connected graphs require many SWAP operations to overcome the limited connectivity. 
Running the same benchmarks on the D-Wave QA requires only 102 qubits (less than 1\% of the qubits). 
Figure~\ref{fig:googledwavecompare} shows the value of the solution obtained from both machines.

\begin{figure}[ht!]
	\centering
	\includegraphics[width=\columnwidth]{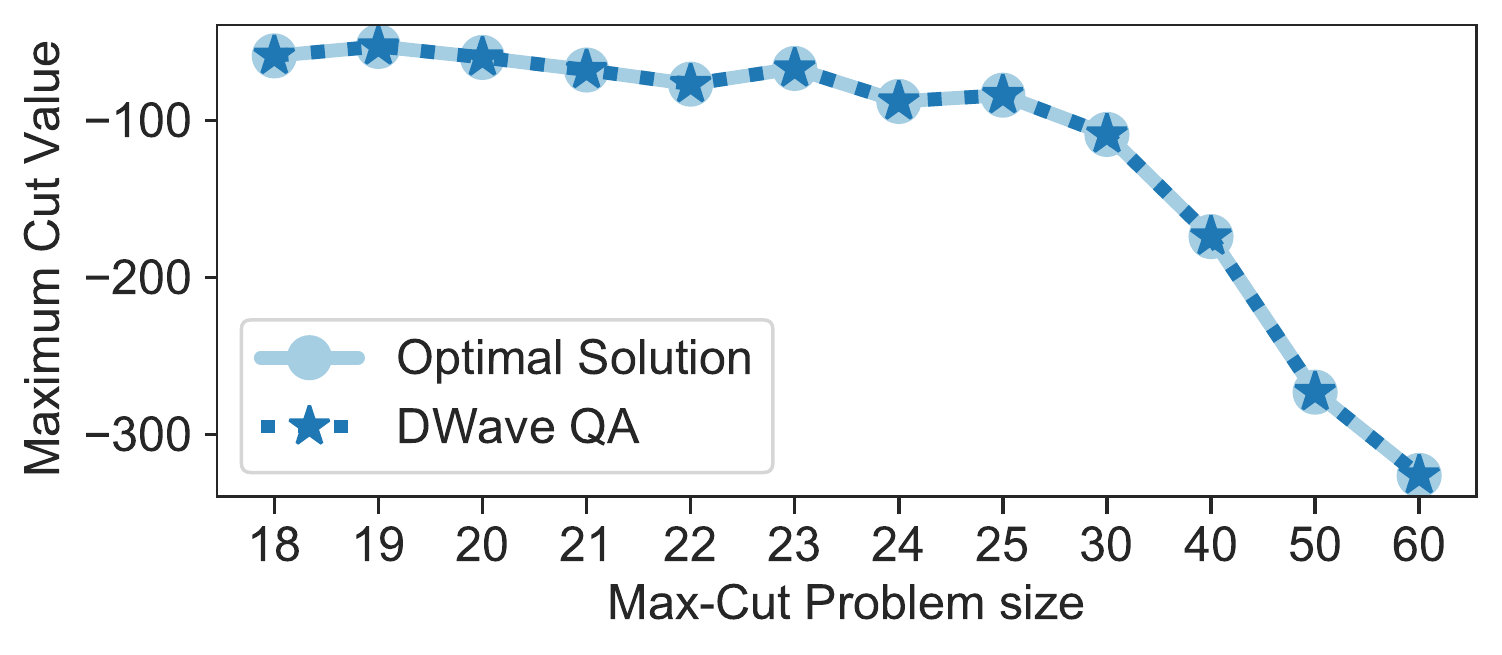}
	
	\caption{Performance of D-Wave QA for larger Max-Cut problems corresponding to the SK Model}
	\label{fig:largeproblemsdwave}
\end{figure}

We use the same weighted graphs from prior work~\cite{harrigan2021quantum} that result in negative cut values. 
The performance of the D-Wave QA is comparable to Google Sycamore and both of them are successful in finding the optimal cut at small problem sizes. 
However, the performance of Google Sycamore degrades with increasing problem size~\cite{harrigan2021quantum} due to an increase in SWAPs and circuit depth. 
Furthermore, due to the limited capacity of 54-qubits, the problem size for Sycamore is limited to no more than 54 nodes. 
However, as QAs are much larger (2000--5000 qubits), we can use them to solve larger problems more relevant to real-world applications and exceed the size of near-term gate-based quantum computers. 
For example, Figure~\ref{fig:largeproblemsdwave} shows the performance of the D-Wave QA for Max-Cut problems corresponding to the SK Model using up to 60 qubits. 
For each problem, the D-Wave QA can find the optimal cut value. 
To determine the optimal cut value, we evaluate all possible combinations for problems using up to 25 qubits and use the best estimate for the larger problems.



\subsection{The Challenge: Hardware and Software Limitations}
Although QAs look promising for various applications, several challenges limit us from solving real-world problems on them.

\vspace{0.05 in}
\subsubsection{Hardware-Level Challenges} \hfill \\

\noindent \textbf{Limited coherence/annealing time:} 
The probability of finding the ground state of a Hamiltonian using a QMI increases exponentially with increasing annealing time~\cite{albash2018adiabatic} and theoretically, many hard problems may require large annealing time. 
Unfortunately, the annealing time on current QAs is in the order of microseconds~\cite{mcgeoch2020theory, ayanzadeh2020multi} as qubits can retain their state only for a short span of time. 
Increasing the annealing time on QA hardware causes qubits to \emph{decohere} and lose their state.

\vspace{0.05 in}
\noindent \textbf{Noise and limited connectivity:}
Thermal noise and operational errors add unwanted perturbations during annealing and prevent QAs from reaching the ground state of a Hamiltonian~\cite{albash2018adiabatic}. 
QAs also suffer from sparse connectivity between qubits, as shown in Figure~\ref{fig:DW_Chimera}. 
To address the same drawback on gate-model quantum computers, compilers insert SWAP instructions that interchange the state of physically adjacent qubits~\cite{tannu2019not,zulehner2018efficient,murali2019noise}. 
However, QAs cannot use a similar approach as they use only one QMI. 
Instead, we \emph{embed} the problem graph to match the target device topology where multiple physical qubits represent a program qubit with higher connectivity. 
This can reduce the effective capacity of QAs.  

\vspace{0.05 in}
\noindent \textbf{Limited precision and range of coefficients:} 
Casting a problem to a Hamiltonian and generating the corresponding QMI coefficients can require a double-precision representation. 
However, large precision impacts the performance of the digital-to-analog converters (DACs) used on the real QAs which slows the annealing process. 
Therefore, existing QAs trade-off precision to achieve lower annealing times and truncate the QMI coefficients post casting to match the precision supported in hardware. 
This subjects the QMI to quantization errors, and the reduced precision QMI actually executed on a QA can be slightly different from the QMI that we originally intended to run, leading to a ground state that may not represent the solution of the problem at hand~\cite{dorband2018extending,pudenz2015quantum,ayanzadeh2020multi}.

\vspace{0.05 in}
\subsubsection{Software-level Challenges} \hfill \\
\noindent \textbf{Limited programmability:} 
QA can only minimize a specific objective function and any input problem must be \emph{cast} to a Hamiltonian. 
Unfortunately, casting is non-trivial due to lack of standardized algorithms and often comes with some approximations~\cite{mcgeoch2020theory,ayanzadeh2020reinforcement}. Additionally, QAs can only execute a single QMI that performs the annealing step and therefore, fine-grained optimizations at the instruction-level are infeasible.

\vspace{0.05 in}
\noindent \textbf{Limitations of Embedding:} 
To overcome limited connectivity, a problem graph is embedded in the QA to match the device topology. 
Finding the best embedding is NP-hard~\cite{cai2014practical,date2019efficiently,goodrich2018optimizing,boothby2016fast} 
and existing algorithms can take several hours despite approximations. 
Also, our studies show they often fail for large programs.

\subsection{Impact of Trials on Energy Residual}
In theory, QAs should find the global optima of a problem with high probability~\cite{nishimori2017exponential,albash2018adiabatic}. 
However, in reality, QAs often fail to find the global optima for large problems due to noise and imperfect control. 
Moreover, the limited programmability of QAs forces users to run a single QMI for thousands of trials, resulting in a bias. 
As a user runs a single QMI for all trials, the noise profile is similar throughout execution, resulting in similar quality outcomes due to the inherent bias in the noise profile. 
We refer to this bias as \emph{Systematic Bias}.  

\begin{figure}[ht!]
	\centering
	\includegraphics[width=\columnwidth]{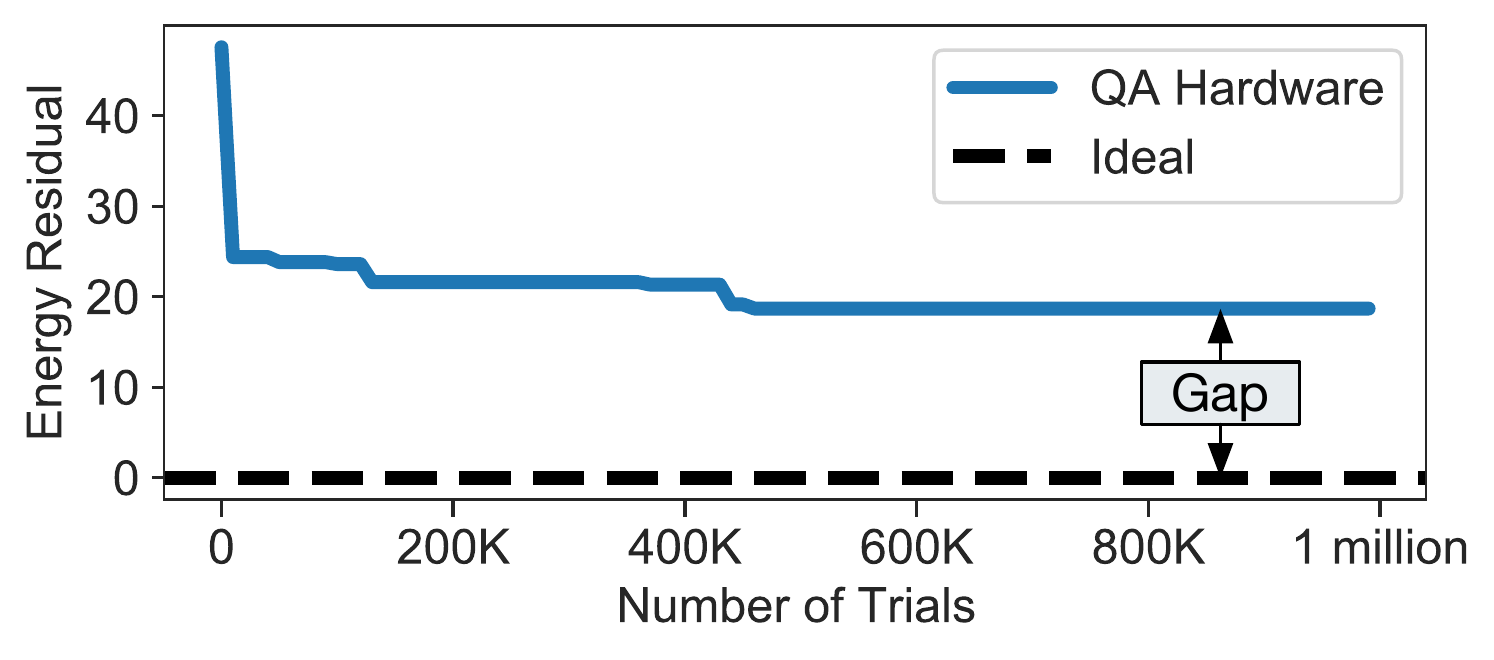}
	\caption{Energy Residual of an optimization problem on D-Wave QA with increasing number of trials. 
	}
	\label{fig:motivation}
\end{figure}

Figure~\ref{fig:motivation} shows the \emph{Energy Residual (ER)}~\cite{karimi2017effective,karimi2017boosting} for an optimization problem on D-Wave QA. 
ER compares the gap between the energy of the  solution from a noisy QA and the global minima.  
The energy of the best solution from a noisy QA remains far from the global optima even after running 1 million trials. 
This non-zero ER occurs due to systematic bias, and is particularly severe for large problems. 


\subsection{Goal of this Paper}
Hardware and software limitations of QAs cause programs to encounter a systematic bias during execution which cannot be bypassed by executing more trials. 
Ideally, we want QAs to be free from systematic bias. 
In this paper, we propose Ensemble Quantum Annealing (EQUAL) that uses an ensemble of QMIs (with different biases) to mitigate systematic bias. 
We discuss the evaluation methodology before discussing our solution.

\section{Evaluation Methodology}
We discuss the evaluation infrastructure used in this paper.

\subsection{Quantum Platform and Baseline}
For our evaluations, we use the 2041-qubit quantum annealer from D-Wave Systems via Amazon BraKet cloud service~\cite{AmazonBraKet}. 
We use the default annealing time (i.e., 20 $\mu$seconds) and schedule recommended for this system.
For the baseline, we use 100,000 trials for each benchmark. 
For EQUAL, trials are equally split between QMIs. 
Thus, EQUAL requires the same number of trials as the baseline.

\subsection{Benchmarks}
We use \emph{random weighted Max-Cut} problems, similar to Quantum Approximate Optimization Algorithms~\cite{farhi2014quantum} used on gate-based quantum computers. %
For the benchmarks, we draw the Hamiltonian coefficients of the QMIs from the standard normal distribution (a mean of 0 and standard deviation of 1). 
This approach is a common practice used in prior works related to benchmarking QAs~\cite{das2008colloquium,pudenz2015quantum,ayanzadeh2020multi,borle2019post,ayanzadeh2019quantum_assisted}. 
To avoid the impact of embedding on our evaluations, we directly use the connectivity graph of the D-Wave QA. 
Thus, the number of program qubits in benchmarks is equal to the number of physical qubits on the QA. 
As the size of benchmarks significantly exceeds the size of existing gate-model quantum computers, we cannot compare our results with them.

\subsection{Figure-of-Merit}
We evaluate the reliability of QA using \textbf{Energy Residual (ER)}. 
The best solution from a QA is the outcome with the minimum energy. 
ER computes the energy gap between the minimum energy $(E_{min})$ obtained on a QA with respect to the global energy minimum $(E_{global})$ of the application as follows:
\begin{equation}
    \label{eq:eres}
    \textrm{Energy Residual (ER)} = \left| E_{min} - E_{global} \right|. 
\end{equation}
Ideally, when the best solution obtained on a QA corresponds to the ground state of the problem Hamiltonian, ER is zero. Thus, a lower value (closer to zero) for ER is desirable.

The challenge in computing the ER for random large benchmarks spanning 2000+ qubits is that finding the ground state of the Hamiltonian is non-trivial. 
To overcome this challenge and still enable a fair comparison, we perform intensive classical computations using state-of-the-art tools~\cite{ayanzadeh_ramin_2021_5142230} and approximate the global optimum of our benchmark problems. 
Recent studies have shown that this algorithm can estimate the ground state of Chimera based Hamiltonians~\cite{cai2014practical,D-Wave} (such as the ones considered in our paper) with a very high probability. \footnote{The techniques used to derive the best estimate of the ground state energy of a Hamiltonian requires intensive classical computing resources and could take up-to days for problem sizes with a few thousand qubits. We discuss more on this in Section~\ref{sec:related_work}}

\section{EQUAL: Ensemble Quantum Annealing}
The vulnerability of a program to systematic bias results from limited programmability and the current execution model of QAs where the same QMI is executed for thousands of trials. 
This subjects each trial to a similar noise profile on the QA and the entire execution suffers from the same inherent bias. 
Our proposed solution \emph{EQUAL}---Ensemble Quantum Annealing---takes a different approach. 
Instead of a single QMI, EQUAL generates an ensemble of QMIs that subjects the program execution to different noise profiles and therefore, different systematic biases. 
When results are aggregated, ensembles enable us to improve the quality of solutions.

\begin{figure*}[ht]	
	\centering
	\includegraphics[width=6.5in]{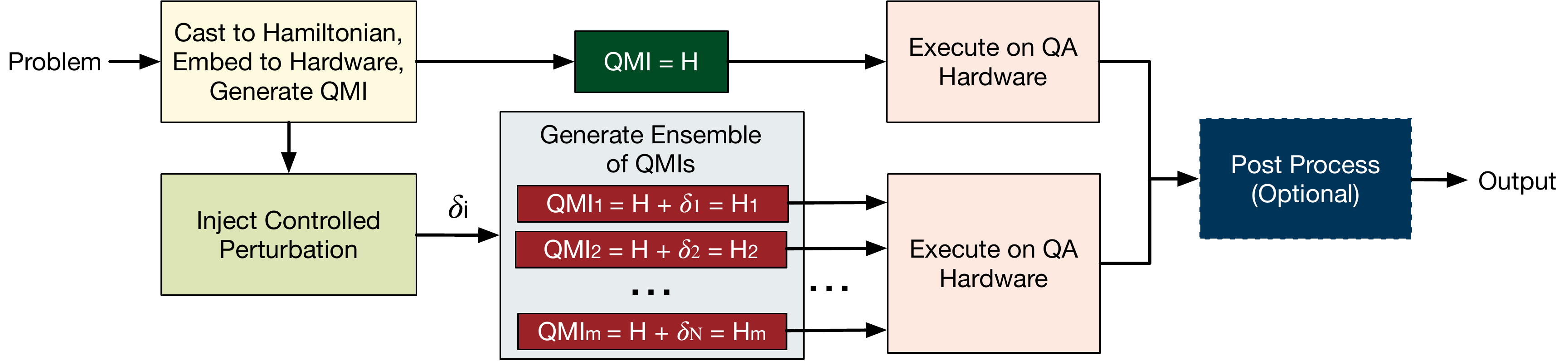}
    \caption{Overview of EQUAL. EQUAL creates an ensemble of QMIs by adding controlled perturbations to the original QMI. It executes the original QMI as well as the ensemble of QMIs separately on the QA hardware and returns the outcome with the lowest energy value. EQUAL can also optionally leverage existing post-processing error mitigation schemes (EQUAL+). 
	}
    \vspace{-0.25in}
	\label{fig:overview}
	\vspace{1.0cm}
\end{figure*}

\subsection{Challenges in Generating Ensembles in EQUAL}
There is potential to generate ensembles during any one of the three phases that a problem goes through before execution on a physical QA hardware: (1) casting, (2) embedding, and (3) QMI generation. 
Generating ensembles during casting was previously studied in the context of Boolean satisfiability (SAT)~\cite{ayanzadeh2020reinforcement} and binary compressive sensing~\cite{ayanzadeh2020ensemble} problems on QAs. 
Unfortunately, these methods exploit the features of the application-specific casting algorithms. Therefore, this approach has limited applicability and is hard to generalize for QAs. 
The other alternative approach is to use an ensemble of embeddings for a given problem. However, this approach too has its limitations. 
Firstly, finding the best embedding is an NP-hard problem in itself~\cite{cai2014practical,date2019efficiently,goodrich2018optimizing,boothby2016fast}. 
Secondly, current embedding schemes for QAs use several approximations and may or may not be able to determine an ensemble of embeddings of similar quality~\cite{cai2014practical,goodrich2018optimizing,date2019efficiently,boothby2016fast}. 
Our studies show that existing embedding algorithms often fail to find an adequate number of embeddings, particularly for problems at scale that require 2000+ qubits. 
Thirdly, even if it is possible to find multiple embeddings, they are often of inferior quality and require larger chains of physical qubits to represent a program qubit with higher connectivity. 
This makes the embedding significantly more vulnerable to noise compared to the best embedding. Thus, generating ensembles at the embedding step is non-trivial. Instead, EQUAL focuses on generating ensembles at the instruction-level and produces multiple QMIs.

\subsection{Overview of Design}
Figure~\ref{fig:overview} shows an overview of EQUAL. 
It relies on adding controlled perturbations to the original QMI. 
For each ensemble, EQUAL generates a \emph{Perturbation Hamiltonian}, denoted by $\delta.$   
Each of these Perturbation Hamiltonians creates a new QMI when added to the original Hamiltonian. For example, if EQUAL generates $m$ ensembles of QMIs, it generates $m$ perturbation Hamiltonians, namely $\delta_1, \delta_2, \dots, \delta_m.$  
The ensemble QMIs---QMI$_1$, QMI$_2$ to QMI$_m$---are obtained by adding the original Hamiltonian (say $\mathbf{H})$ and the respective perturbation Hamiltonians. 
In other words, the ensemble of QMIs now corresponds to the perturbed versions of the original Hamiltonian.

\subsection{Generating Ensembles via Controlled Perturbations}
\begin{figure}[ht!]
	\centering
	\includegraphics[width=0.9\columnwidth]{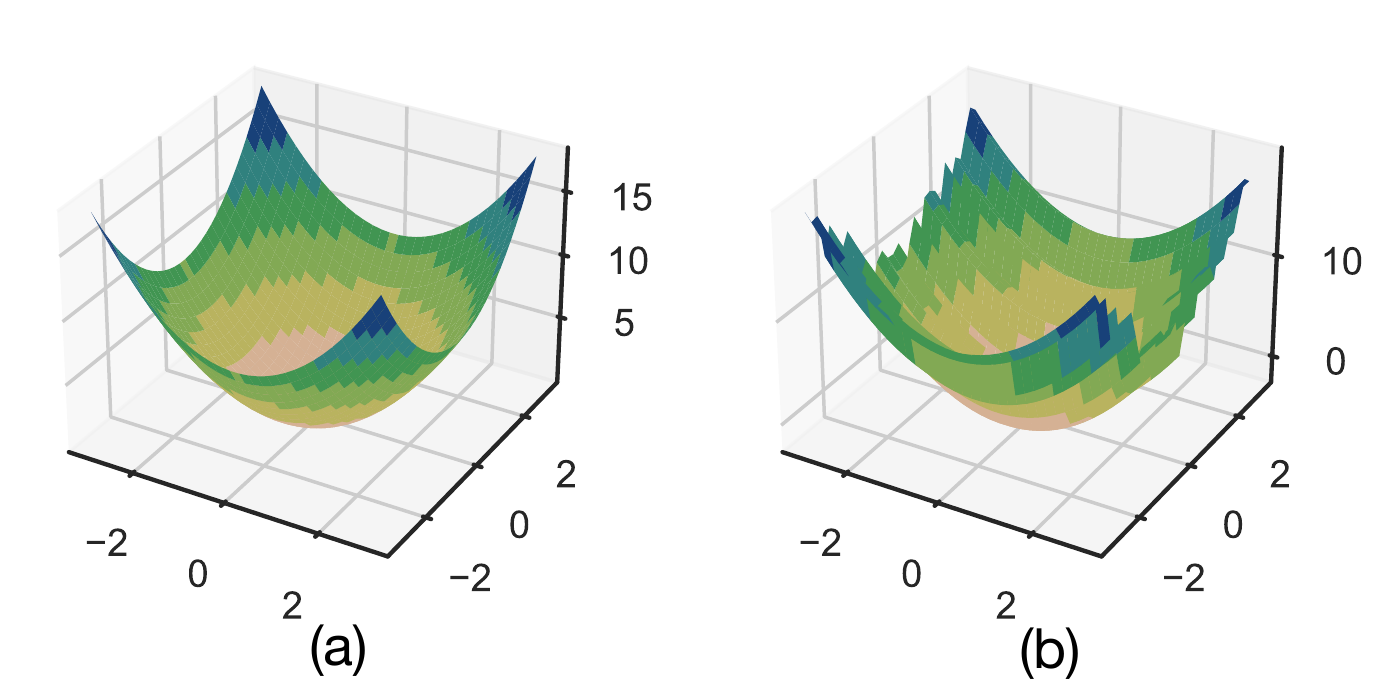}
	\caption{
		(a) Landscape of an example optimization problem. 
		(b) The resultant landscape differs significantly from (a) when an extremely noisy perturbation is imposed. (This figure is for illustrative purposes only. Hamiltonians and QAs can only deal with discrete optimization problems.)
	}
	\label{fig:landscape}
\end{figure}

Creating an effective perturbation Hamiltonian is non-trivial. 
If the perturbations add too little noise, the resulting Hamiltonian will be too close to the problem Hamiltonian and encounter similar bias. 
Alternately, too large perturbations result in a Hamiltonian significantly different from the problem of interest and can produce infeasible results.  
For example, Figure~\ref{fig:landscape}(a) shows the landscape of an example optimization problem\footnote{Minimize $x^2+y^2$ for $x,y \in [-2,2]$.}. 
Figure~\ref{fig:landscape}(b) shows that injecting an extremely noisy perturbation Hamiltonian significantly changes the landscape of the original problem.  
Thus, there is a trade-off between the effectiveness of a perturbation Hamiltonian to reduce bias and its ability to alter the problem Hamiltonian. 
To address this challenge and generate an effective ensemble of QMIs, EQUAL exploits the device-level characteristics of QAs. 

\vspace{0.05in}
\subsubsection{Exploiting Hardware Characteristics of QAs} \hfill

\noindent Recollect that casting a problem to a Hamiltonian can require double-precision representation of the Hamiltonian coefficients. 
Unfortunately, real QAs can only support a small range and precision of coefficients due to the limitations imposed by the digital to analog converters (DACs) used on QAs. If the precision of the coefficients are too large, the DACs are too slow which eventually slows the controlling modules of QAs and is not desirable. 
To bridge this gap, post the casting step, the coefficients of the QMI are truncated to match the precision supported by the hardware. 
While this is a limitation on QAs, EQUAL leverages it to its advantage and draws the coefficients of the perturbation Hamiltonian randomly at a range that is below the supported precision so that adding the perturbation Hamiltonian only shifts the coefficients of the QMI (post truncation) to one of the neighboring quantization levels and thus, does not significantly alter the problem landscape. 
More specifically, let $b$ be the number of bits used for representing coefficients of a physical QA. 
For every ensemble, EQUAL draws a uniform random number 
$r \in \left[{\frac{1}{2^{b+1}}, \frac{1}{2^b}}\right]$
and set all coefficients of the Perturbation Hamiltonian to be $r$.

\begin{figure*}[b!]	
\vspace{0.15in}
	\centering
	\includegraphics[width=0.95\linewidth]{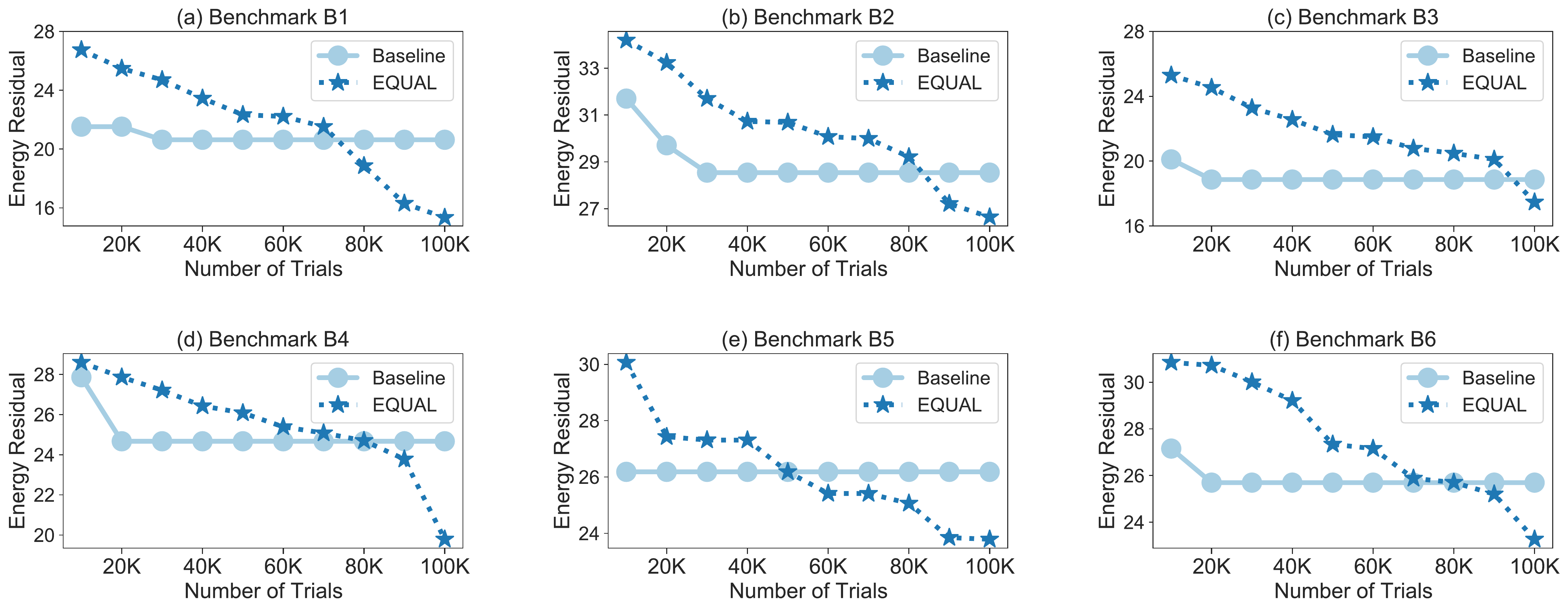}
    \caption{Trends in Energy Residual for the baseline and EQUAL for all the benchmarks.
    \vspace{0.1 in}
	}
	\label{fig:energyresidual}
\end{figure*}  

\subsubsection{Profiling QAs to estimate Hardware Precision} \hfill

\begin{figure}[ht]
	\centering
	\includegraphics[width=\columnwidth]{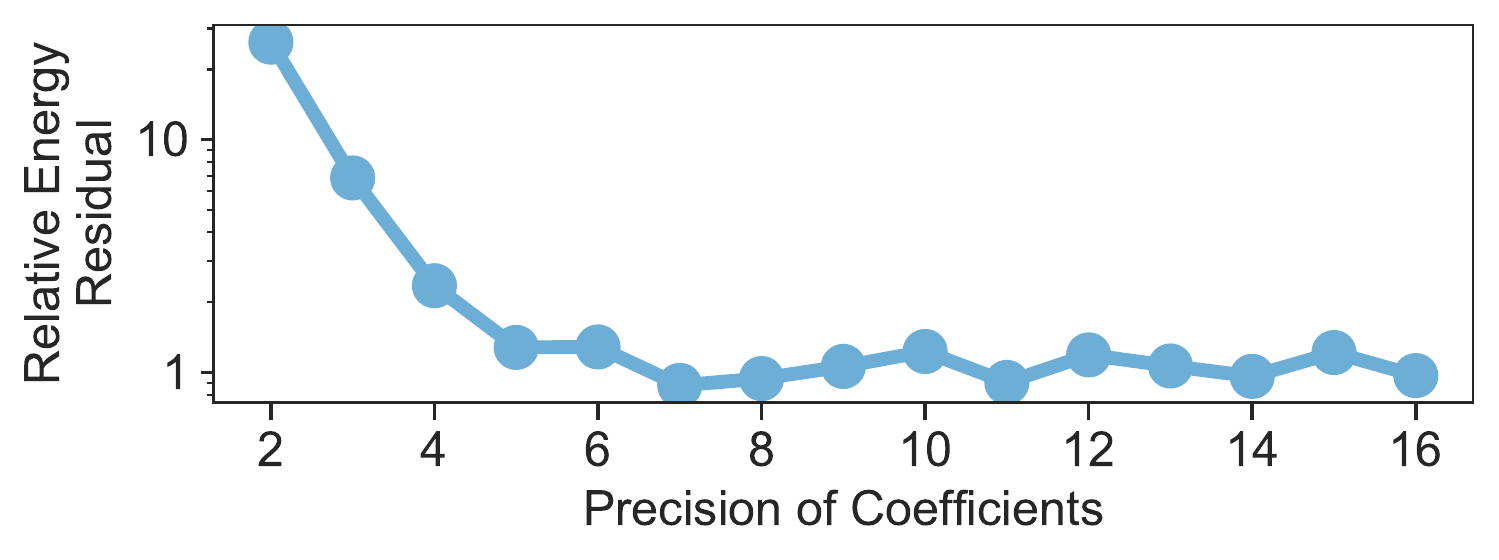}
	\caption{
	Relative Energy Residual of QMIs with truncated coefficients with respect to the original problem QMI for bits values of precision. (Lower is better)
	}
	\label{fig:precision_rDivergence}
\end{figure}

\noindent Unfortunately, the precision of the coefficients supported on real devices is unavailable to programmers. 
Determining this precision is vital for the performance of EQUAL. 
Drawing the perturbation Hamiltonian coefficients far below the supported precision introduces large noise and may alter the Hamiltonian landscape significantly. 
Alternately, drawing them far above the supported range may not have any effect post truncation. 
To tackle this challenge, EQUAL profiles the QA using random benchmarks to estimate the precision supported by QAs. 
In this experiment, we truncate all coefficients of the benchmark for $2, 3, \dots, 16$ bits precision and execute the corresponding QMIs. 
Figure \ref{fig:precision_rDivergence} shows the relative Energy Residual of the truncated QMIs with respect to the original problem (without truncation). 
Our profiling experiments with multiple benchmarks show that the hardware is likely limited by 7--8 bits of precision.  
Thus, EQUAL generates coefficients of ensembles in $\left[\frac{1}{2^9}, \frac{1}{2^8} \right].$ 

\subsection{Execution on QA Hardware}
EQUAL splits the trials between the ensemble of QMIs equally, including the original QMI (without perturbation), and executes them separately on the QA hardware. 
Our default design uses 10 ensembles of QMIs, and allocates 10,000 trials for every ensemble. 
We do a more rigorous sensitivity analysis for the number of trials and ensembles in Section~\ref{sec:sensitivity}.

By default, the outcome with lowest energy is deemed as the solution for problems executed on QAs. 
In the baseline, this corresponds to the outcome with the lowest energy obtained by executing the original QMI. 
As EQUAL executes multiple QMIs, the outcome with the lowest energy among all the QMIs is returned as the solution. 
Also, as EQUAL runs the ensemble of perturbed QMIs in addition to the original program QMI, the final solution is guaranteed not to perform worse than the baseline, assuming there are no sampling errors. 
Note that the solution with the minimum energy corresponds to an outcome that may come from a single QMI. 
For the baseline this corresponds to the original QMI, whereas for EQUAL it comes from one or more of the QMIs in the ensemble. 
However, which QMI corresponds to the best solution is not known a-priori and EQUAL must execute the entire ensemble.

\begin{figure*}[ht]
	\centering
	\includegraphics[width=1\linewidth]{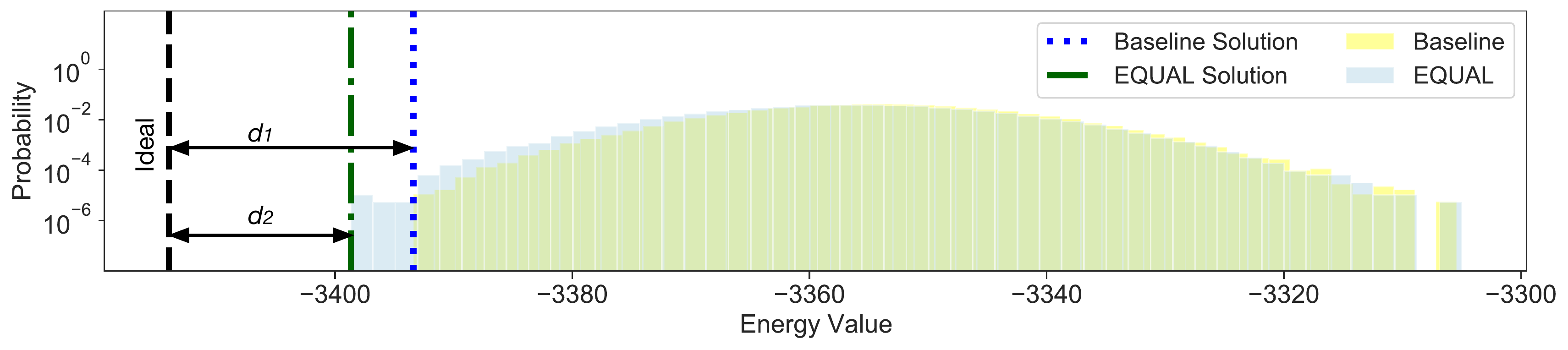}
	\caption{
	Histogram of energy values from the outcomes on the QA for benchmark B1 using the Baseline and EQUAL. The solution from EQUAL is closer to the ideal solution compared to the baseline solution ($d_2<d_1$). The histograms for the baseline and EQUAL largely overlaps which indicates that EQUAL does not significantly alter the problem Hamiltonian.
	\vspace{0.1 in}
	}
	\label{fig:equal_histogram}
\end{figure*}

\subsection{Results for Energy Residual}
Energy Residual (ER) computes the gap between the energy obtained from the best outcome on a QA with the global optima. Figure~\ref{fig:energyresidual} compares the ER of the individual benchmarks for baseline and EQUAL. 
We observe that the ER quickly saturates in the baseline for all benchmarks, whereas improves for EQUAL as more QMIs are executed. 
As the QMIs are generated using random controlled perturbations, some of them may result in higher ER compared to the baseline due to a different noise profile at run time. 
However, the ensemble overall enables EQUAL to reach a better solution. 
In the worst case, EQUAL performs similar to baseline as the original program QMI is executed too. 
We observe that the fidelity of the baseline saturates with more trials, whereas the diversity of EQUAL helps it keep on improving with additional trials.

\begin{figure}[htp]
	\centering
	\includegraphics[width=\columnwidth]{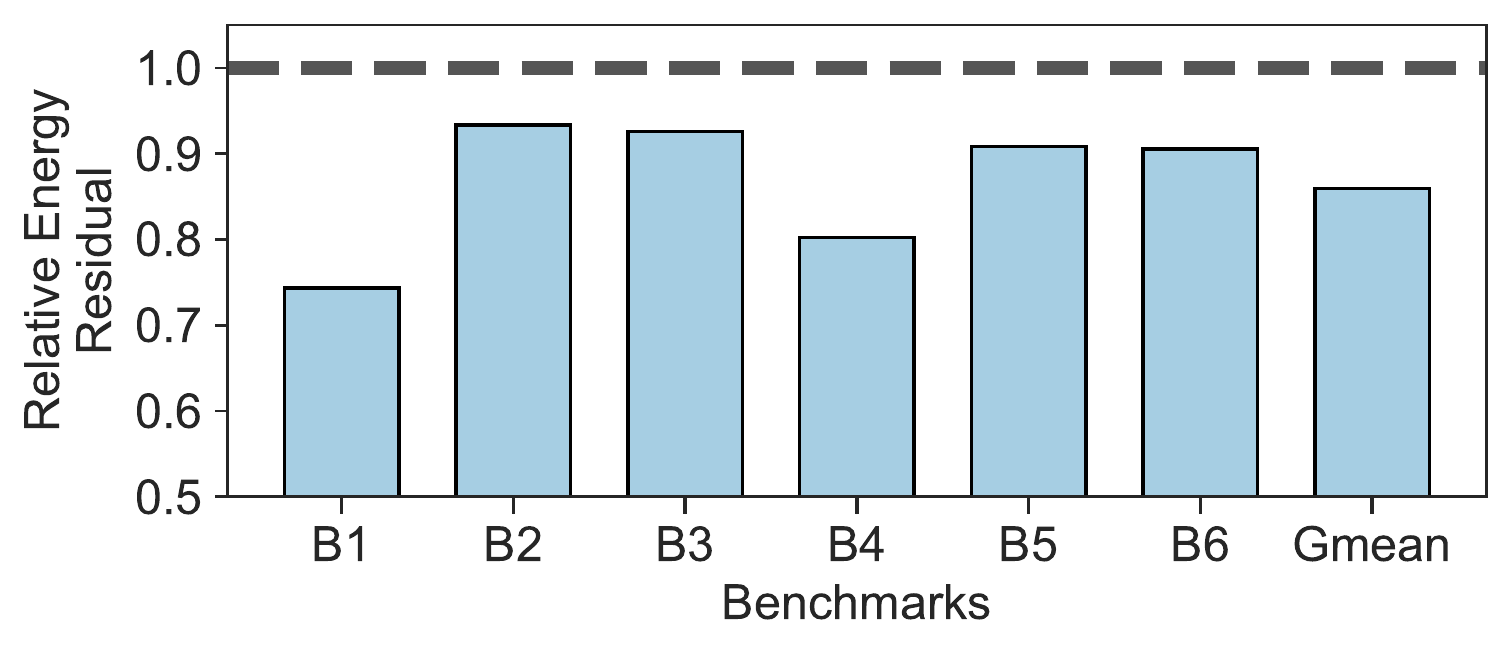}
	\caption{Energy Residual of six random benchmarks on D-Wave QA hardware using EQUAL relative to the baseline.
	}
	\label{fig:eval}
\end{figure}

Figure~\ref{fig:eval} shows the ER of EQUAL for our benchmarks executed on D-Wave 2041-qubit QA relative to the baseline. 
We observe that EQUAL bridges the  difference between the baseline and the ideal by an average of 14\% (and up to 26\%). 
QAs deal with industry-scale optimization problems where even a minuscule improvement has a tremendous impact in terms of practical advantage such as saving millions of dollars~\cite{ahuja2005network,carlson2012miso,elsokkary2017financial} in the context of scheduling and planning applications or finding better candidates for drug discovery~\cite{mulligan2020designing} and material science~\cite{kitai2020designing}.  
Thus, the quality of solutions is of utmost importance.  

\subsection{Case-Study: How EQUAL Reduces Systematic Bias}
Figure~\ref{fig:equal_histogram} shows the histograms of the energy values obtained by running benchmark B1 for the baseline and EQUAL. 
The goal of QAs is to obtain the outcome corresponding to the ground state energy. 
We observe that the optimal solution is at a distance $d_1$ from the ground state and EQUAL produces a solution at distance $d_2$ that is closer to the ground state energy ($d_2 < d_1$) by minimizing the impact of bias. 
We also observe that distributions for both the baseline and EQUAL overlap largely, indicating that the ensemble of QMIs do not largely alter the original Hamiltonian corresponding to our problem. We make similar observations for other benchmarks.

An ideal QA is a machine that samples from the Boltzmann distribution, whereas the distribution obtained from a real QA hardware is different due to noise. 
The best solution obtained by a QA depends on the overlapping region between the ideal and the noisy distributions. 
From Figure~\ref{fig:equal_histogram}, we observe two potential approaches to get closer to the global optima. 
First, by \emph{flattening} the energy histogram of the Hamiltonian such that it covers a broader search space. 
Second, by \emph{shifting} the energy histogram towards the ideal solution. 
Note that both of these techniques must ensure that the properties of the original program Hamiltonian remain unaltered. 
EQUAL uses the first approach. 
The performance of EQUAL can be improved further if we could shift the histogram closer to the ideal solution. 
We explore combining EQUAL with existing error mitigation schemes to obtain the advantage of both flattening the histogram and shifting the histogram towards the ground state.

\section{Combining EQUAL with Error-Mitigation}

Ensembles are generated by only adding controlled perturbations to the problem Hamiltonian. 
Therefore, they have limited capability to shift the noisy distribution from a QA towards the ideal distribution even if a large number of ensembles are used. 
Alternately, large perturbations may significantly change the landscape of the problem. 
Instead, we take an orthogonal approach and explore existing error-mitigation schemes that introduce a shift in the energy histogram.

\subsection{Primer on Error-Mitigation Schemes for QA}
Error-mitigation schemes for QAs can be classified into (1) software and (2) hardware schemes. 
Software schemes refer to optimizations performed during the casting and embedding steps (pre-processing techniques) or modifications on the outcomes obtained from QAs (post-processing techniques). 
On the other hand, hardware-based schemes control the device-level parameters on QAs to reduce the impact of errors. 

We characterize the impact of these error-mitigation techniques individually and combined with each other to understand their effectiveness in (1)~eliminating systematic bias on their own and (2)~shifting the noisy distribution of the QA towards the ideal distribution. 
For our analysis, we choose (a)~spin reversal transform, (b)~longer inter-sample delay, and (c)~single-qubit correction. 
Spin-reversal transform is a representative candidate for a software pre-processing technique. 
On the other hand, inter-sample delay is a device-level control available to programmers to reduce the correlation between consecutive trials on a QA. 
Lastly, Single-Qubit Correction (SQC)~\cite{ayanzadeh2020multi} is a post-processing technique which leverages the insight that for a given QMI, QAs can quickly recognize the neighborhood of the ground state even if they fail to get to the ground state~\cite{ayanzadeh2020multi}. 
We perform characterization studies for these three error mitigation schemes (see Appendix~\ref{sec:additionalchar}) and found that SQC is the most effective scheme, and therefore we use SQC as the error mitigation scheme for our study.

\begin{figure}[htb]
	\centering
	\includegraphics[width=\columnwidth]{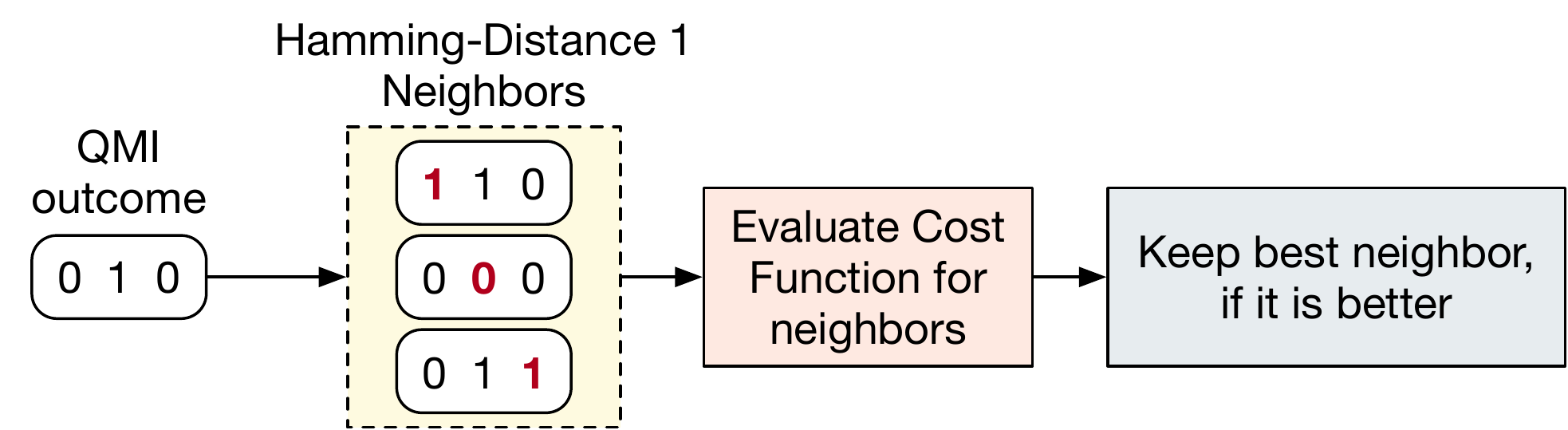}
	\caption{Single Qubit Correction Post-Processing~\cite{ayanzadeh2020multi}
	}
	\label{fig:sqc_block_diagram}
\end{figure}  

\subsection{Overview of SQC Post-Processing}
SQC is analogous to the gradient descent scheme but only applicable to discrete optimization problems. Instead of computing the gradient for determining the direction of the move in every iteration, SQC uses a greedy approach and moves to a neighbor (i.e., an outcome that is one Hamming distance away from the current solution) with the lowest energy value. Figure~\ref{fig:sqc_block_diagram} illustrates the overview of an iteration of SQC. For each candidate outcome generated by executing the QMI, SQC finds the one Hamming distance away neighbors and computes their energy values. If any neighbors can obtain a lower energy value than the candidate itself, the neighbor is retained and the current solution is discarded. When multiple neighbors obtain lower energy values, the best neighbor is retained. The process is repeatedly executed until we cannot find any new neighbor that has better quality.

 \subsection{EQUAL+: Combining EQUAL and SQC}
\begin{figure}[htp]
	\centering
	\includegraphics[width=\columnwidth]{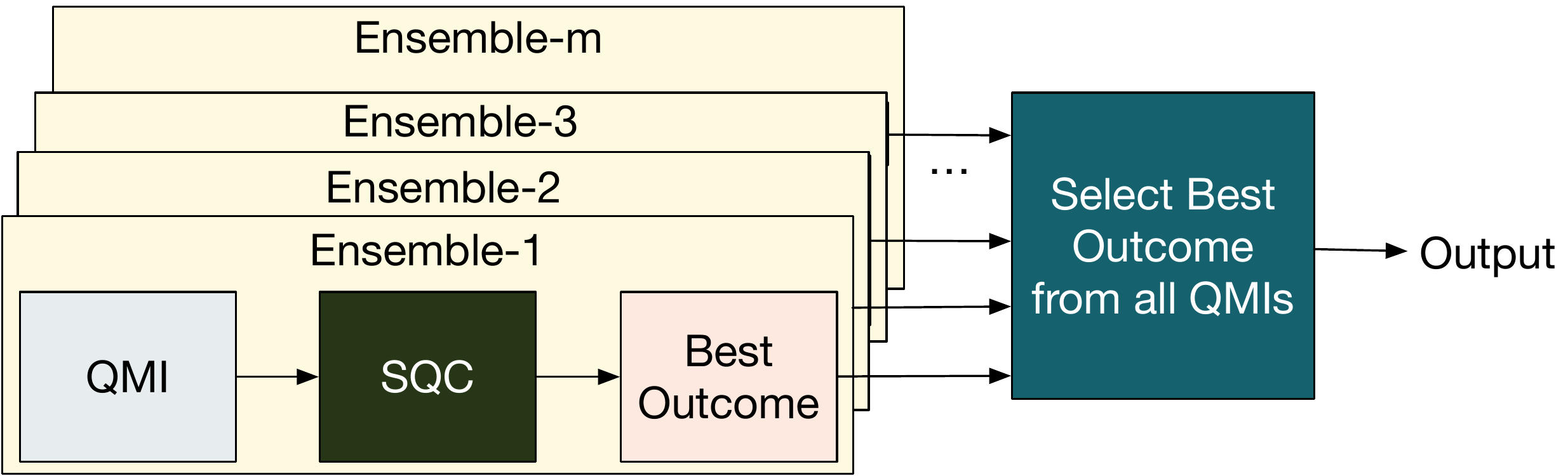}
	\caption{Overview of EQUAL+ design. It applies the SQC post-processing algorithm on the outcomes from each QMI in parallel. Finally, it selects the best outcome from all the QMIs as the output solution.
	}
	\label{fig:block_diag_eqp}
\end{figure}

Figure~\ref{fig:block_diag_eqp} shows an overview of EQUAL+. EQUAL+ applies SQC on the outcomes of each QMI and obtains the best outcome for each QMI. The process is performed for each QMI in parallel. Once applying SQC on each QMI converges, the final output of EQUAL+ is picked as the candidate with the lowest energy among all the individual best candidates from the QMIs. The time to converge depends on several factors such as the size of the problem, number of outcomes, quality of the outcomes. However, our evaluations show that EQUAL+ converges within a few seconds even for large benchmarks such as the ones used in our evaluations.

Using this greedy approach helps locate neighbors from current outcomes that were not produced by the QA originally. With each neighbor located, EQUAL+ shifts the outcome distribution towards the ideal solution (global optima). 
Note that although SQC is effective on its own, the diversity of EQUAL+ is essential to improve its search space. The capability of SQC alone to introduce new outcomes is limited by the quality of outcomes from the QMI. In EQUAL+, the ensembles enable us to explore a much larger neighborhood compared to applying SQC alone. In the end, EQUAL+ may discover a solution from one of the weakest outcomes corresponding to one of the weakest QMIs (sub-optimal outcome that did not correspond to the best solution in any of the QMIs). 

Note that EQUAL+ is versatile, and any other post-processing candidate that introduces the desired shifting property in the energy distribution may be used. We use SQC for its performance and low time complexity.


 \begin{figure*}[htb]
	\centering
	\includegraphics[width=0.9\linewidth]{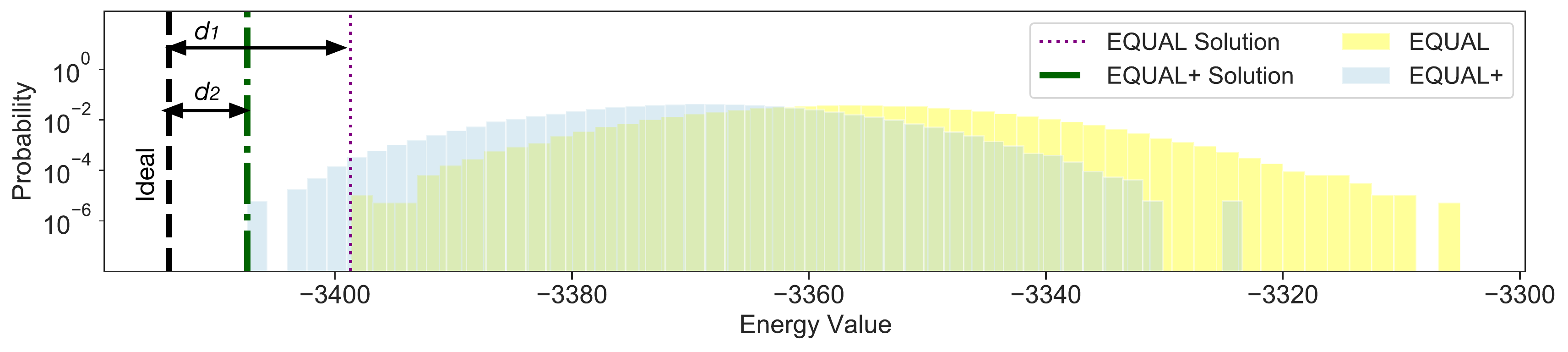}
	\caption{Histogram of energy values from the outcomes on the QA for benchmark B1 using EQUAL and EQUAL+.
	}
	\label{fig:energyhistogram}
\end{figure*}  

\subsection{Analysis of Overheads}
We discuss the overheads for both EQUAL and EQUAL+.
EQUAL generates the ensemble of QMIs prior to execution on the QA. As the perturbed Hamiltonian only adjusts the coefficients of the original problem QMI, the ensemble does not need to re-perform the casting or embedding step. Although embedding can take up to several hours and may fail for certain Hamiltonians, this overhead and limitation is entirely avoided by EQUAL. EQUAL also requires the programmer to estimate the precision of the hardware using a set of profiling experiments. However, profiling need not be done for each application. As the precision supported is only device-specific, profiling once for each QA hardware is enough and the same information can be re-used for multiple applications. For execution on the QA, EQUAL requires the same number of trials as the baseline and therefore, does not incur any overhead of additional trials.

EQUAL+ incurs some additional overheads for the post-processing step as it applies the SQC heuristic algorithm on all the outcomes obtained from all the QMIs. The space complexity of the post-processing phase in EQUAL+ is linear with the number of qubits~\cite{ayanzadeh2020multi}. As SQC is iteratively applied on every outcome of a QMI, the time complexity depends on the number of outcomes which is equal to the number of trials in the worst-case (assuming each trial generates a unique outcome). The post-processing for each QMI is done in parallel. Our studies show that EQUAL+ converges within a few iterations and the post-processing step for EQUAL+ only takes a few seconds. Therefore, the overheads are acceptable.  

\label{sec:evaluations}
\subsection{Results for Energy Residual}
\begin{figure}[htb]
	\centering
	\includegraphics[width=\columnwidth]{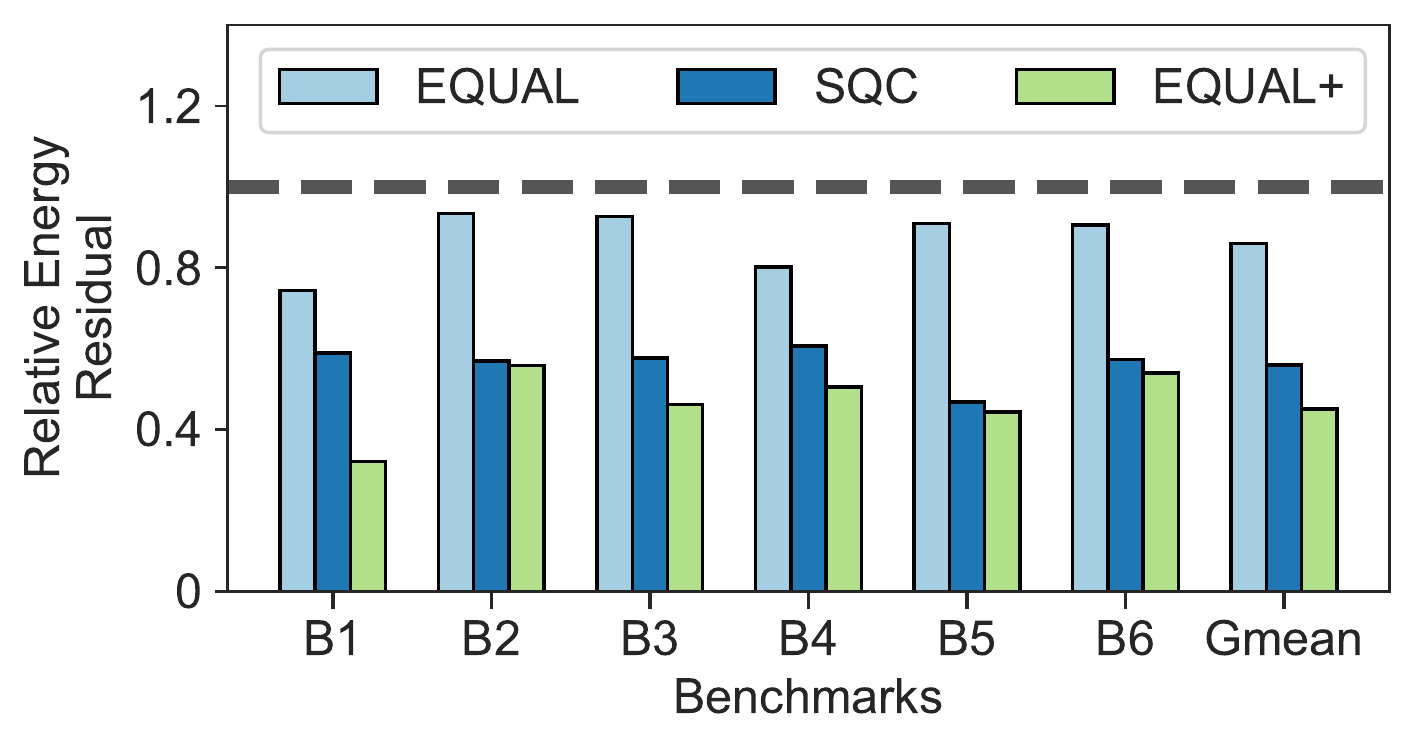}
	\caption{Energy Residual using EQUAL+ relative to the baseline. We also compare with EQUAL and SQC standalone.
	}
	\label{fig:alleval}
\end{figure}

Figure~\ref{fig:alleval} shows the Energy Residual of EQUAL+ relative to the baseline. We also compare against EQUAL and SQC standalone. 
We observe that EQUAL+ improves the ER by 0.45 compared to the baseline on average and by up to 0.32. 
In other words, EQUAL+ improves the quality of solutions by 55\% on average and up to 68\%.

\subsection{Results for Validation of Precision Selection}
\begin{figure}[htp]
	\centering
	\includegraphics[width=\columnwidth]{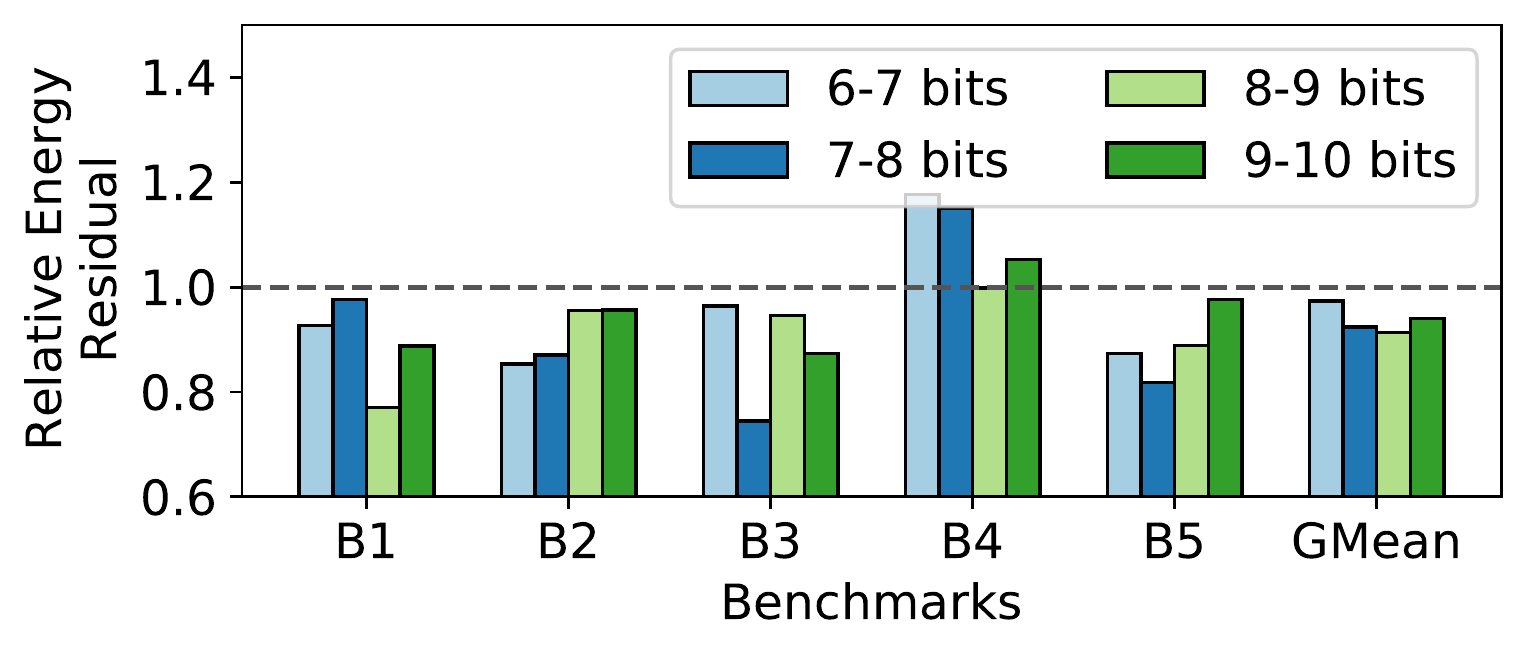}
	\caption{Energy Residual of random benchmarks on D-Wave QA hardware using EQUAL relative to the baseline for different values of bit precision used for generating QMIs.
	}
	\label{fig:precision}
\end{figure}

We draw the perturbation Hamiltonian coefficients in the range $\left[\frac{1}{2^9}, \frac{1}{2^8} \right]$ based on profiling across a wide range of values. We confirm that this approach is robust by conducting additional studies at application level. Figure~\ref{fig:precision} shows the ER of some benchmarks relative to baseline when the precision range is varied. We confirm that 8 to 9 bits of precision is more robust compared to others and on average outperforms the others. 

\subsection{Case-Study: How EQUAL+ reduces Systematic Bias}
Figure~\ref{fig:energyhistogram} shows the histograms of energy values of benchmark B1 for EQUAL and EQUAL+. We observe that the optimal solution is at a distance $d_1$ from the ground state energy in EQUAL. EQUAL+ exploits the shifting property of SQC to obtain a solution at distance $d_2$ and is closer to the ground state energy ($d_2 < d_1$). Note that EQUAL+ shifts the overall histogram towards the ideal solution and achieves the intended goal. As EQUAL+ applies post-processing on the outcomes from the QMIs, the introduced shift in the histogram does not alter the original problem Hamiltonian.

\subsection{Impact of Number of Ensembles}
\label{sec:sensitivity}
\begin{figure}[htp]
	\centering
	\includegraphics[width=\columnwidth]{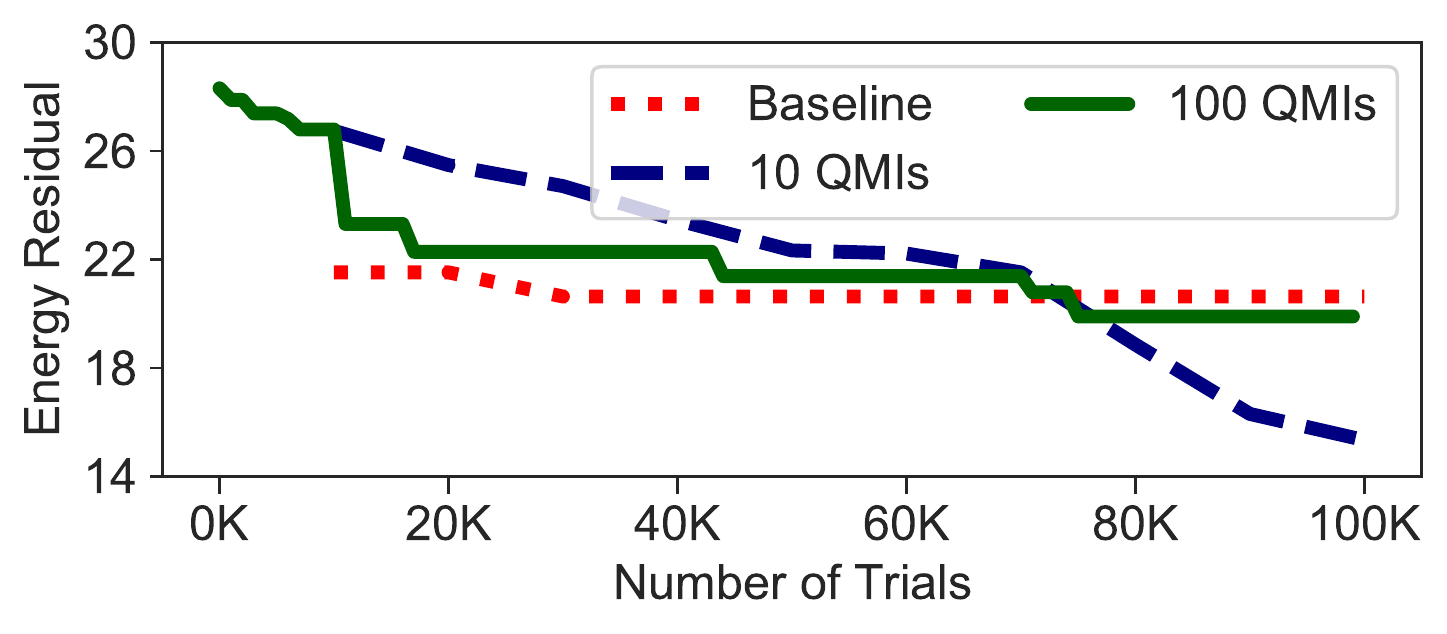}
	\caption{Energy Residual for benchmark (B1). The baseline executes a single QMI for all 100K trials. EQUAL has 10 QMIs for 10K trials each or 100 QMIs executed for 1K trials each. 
	}
	\label{fig:ensembles}
\end{figure}

We study the impact of the number of ensembles on the effectiveness of EQUAL using a single benchmark problem. For a given trial budget of 100K trials, we choose two modes for EQUAL. In the first instance, we use 10 QMIs and run each of them for 10K trials each. In the second instance, we use 100 QMIs and run each of them for 1K trials each. Figure~\ref{fig:ensembles} shows the ER for the baseline and these two instances of EQUAL. Note that we access the QA device through cloud services and a more rigorous sensitivity analysis in terms of QMIs and trials is challenging. 
We observe that executing more QMIs introduces more randomness and makes them vulnerable to sampling errors. EQUAL with 10 QMIs achieve a sweet spot between the baseline and EQUAL with a large number of ensembles such that we have both diversity as well as sufficient trials for each QMI to reduce sampling errors.

\newpage
\section{Related Work}
\label{sec:related_work}
Both gate-model quantum computers and QAs promise significant advantages for a wide range of applications~\cite{arute2019quantum,farhi2014quantum,peruzzo2014variational,kandala2017hardware,biswas2017nasa,rieffel2015case,venturelli2015quantum,tran2016hybrid,bian2016mapping,su2016quantum,ayanzadeh2020reinforcement,o2018nonnegative,peng2019factoring,hu2020quantum,perdomo2015quantum,ayanzadeh2019quantum,ayanzadeh2020ensemble,inoue2021traffic,elsokkary2017financial,kitai2020designing,mulligan2020designing}. 
Thus, developing error-mitigation policies is an active area of research for both QAs and gate-model quantum computers. We discuss prior works and compare against schemes that use ensembles.

\subsection{Priors works using Ensembles}
The potential of ensembles has been explored for both gate-model quantum computers and QAs. 

\vspace{0.05 in}
\noindent \textbf{Ensemble policies for Gate-model quantum computers}:
Systematic bias in QAs is similar to correlated errors on gate-based quantum computers. 
To tackle these errors on gate-model quantum computers, recent studies propose the use of ensemble of mappings that maps a program to different sets of physical qubits and SWAP routes on the same~\cite{tannu2019ensemble} or different machines~\cite{patel2020veritas}. 
This process produces functionally identical copies of the same program but are only executed differently. 
Leveraging a similar approach for EQUAL is non-trivial due to the complexities involved in the embedding process, particularly for problems at scale. Obtaining alternate embeddings is non-trivial, may fail or result in inferior quality. Instead, EQUAL uses an ensemble of QMIs by introducing controlled perturbations while minimizing the alterations in the functionality of the original problem Hamiltonian.

\noindent \textbf{Ensemble policies for QAs}:
Using ensembles in QAs have been investigated at the casting level for two different applications. However, as each application uses its own casting algorithm, this approach cannot be generalized. EQUAL on the other hand, avoids such application-specific assumptions and is applicable irrespective of the problem at hand. In another study, Mohseni et al. proposed a multi-level embedding scheme that uses a diverse encoding of qubits to generate ensembles~\cite{mohseni2021error}. However, this approach reduces the capacity of the QA significantly. It is also not scalable as it introduces overheads to the embedding step, which already takes several hours for current systems. Moreover, our studies show that finding alternate embeddings frequently fail or result in embeddings of inferior quality for large problems. EQUAL avoids these overheads by introducing diversity post embedding.

\subsection{Software error mitigation policies}
These techniques are either applied prior to the execution of the QMI (pre-processing) or after the QMI is executed (post-processing). 
Pre-processing schemes transform the problem QMI at the casting or embedding level such that it is less vulnerable to errors during execution time~\cite{pelofske2019optimizing,ayanzadeh2020multi,ayanzadeh2020reinforcement,ayanzadeh2020ensemble}. 
Pre-processing schemes are analogous to compiler-level optimizations on gate-model quantum 
computers~\cite{tang2021cutqc,murali2019full,patel2020ureqa,li2021large,alam2020efficient,murali2020architecting,murali2020architecting,patel2020disq,shi2019optimized,zulehner2018efficient,barron2020measurement,kwon2020hybrid,gokhale2019partial,tannu2019ensemble,murali2019noise,alam2020circuit,murali2020software,patel2020veritas,tannu2019mitigating,tannu2019not,li2018tackling,gokhale2020optimized}. 
Unlike QAs, gate-model quantum computers have higher programmability and allow programmers to leverage fine-grained optimizations at the instruction scheduling level. 
Post-processing schemes for QAs exploit the fact that even if a QA cannot generate the solution with the lowest energy, it quickly locates the neighborhood where the optimal solution might reside. By modifying the outcome obtained from the QA using classical heuristic algorithms, post-processing schemes can significantly improve the quality of solutions~\cite{ayanzadeh2020multi,borle2019post,golden2019pre}. One of the most promising post-processing schemes is Multi-Qubit Correction (MQC)~\cite{ayanzadeh2020multi}. We use this scheme to obtain the best known estimate of the ground state energy in our evaluations. However, this algorithm requires significant classical computational overheads and may take up to days to obtain a better quality solution. Nonetheless, the performance of MQC depends on the quality of the outcomes obtained from the QA and both EQUAL and EQUAL+ can benefit from it. Post-processing algorithms can significantly improve the application fidelity even for gate-based quantum computers~\cite{matrixmeasurementmitigation,bravyi2020mitigating,patel2020veritas,patel2021qraft}.

\subsection{Hardware-level error mitigation}
Recent studies have proposed improvements to the annealing process itself~\cite{nishimori2017exponential,albash2018adiabatic} but implementing them requires fabricating newer QAs. Few other schemes reduce the impact of noise on the annealing process. Examples include controlling the delay between two consecutive trials and qubit preparation times~\cite{D-Wave_Ocean_SDK}. Exposing the hardware characteristics and enabling device-level controls to the programmer has been shown to be effective even for gate-model systems~\cite{asfaw2019get}.

\section{Conclusion}

Quantum Annealers (QAs) have thousands of qubits and are promising for a wide range of applications. Unfortunately, QA suffers from both hardware and software limitations. Therefore, there is increasing interest in developing software policies to tackle these limitations. However, our studies show that executing a program on a QA makes it vulnerable to a systematic bias that cannot be overcome by increasing the number of trials or relying on existing error-mitigation schemes. 

This paper proposes EQUAL---Ensemble Quantum Annealing---a software framework that creates multiple perturbed copies of an input problem by injecting controlled perturbations to the original problem Hamiltonian. 
By executing an ensemble of quantum machine instructions (QMIs), EQUAL projects the program to different noise profiles and therefore, different biases. 
Our evaluations using the 2041-qubit D-Wave QA show that EQUAL bridges the difference between the baseline and the ideal by an average of 14\% (and up to 26\%), without requiring any additional trials. 
We also propose EQUAL+ that exploits the properties of existing error mitigation schemes for enhanced performance. 
EQUAL+ bridges the difference between the baseline and the ideal by an average of 55\% (and up to 68\%).

\newpage
\appendix

\section{Appendix: Characterizing Error Mitigation}
\label{sec:additionalchar}
While designing EQUAL+, we have various options regarding how to combine existing error-mitigation schemes. To understand suitable candidates, we perform several characterization experiments whose results are described next.

\subsection{Spin-Reversal vs. Single Qubit Correction}

Spin-Reversal Transform (SRT) is a pre-processing which flips randomly selected qubits to mitigate analog errors of QAs~\cite{pelofske2019optimizing}.
Figure~\ref{fig:impactofisd} shows the Energy Residual of a benchmark executed on D-Wave QA when spin reversal transform is applied standalone and in the presence of SQC. We observe that both spin reversal and SQC reduces the ER, and the performance of SQC is comparable to spin-reversal transform. 

\begin{figure}[ht!]
	\centering
	\includegraphics[width=\columnwidth]{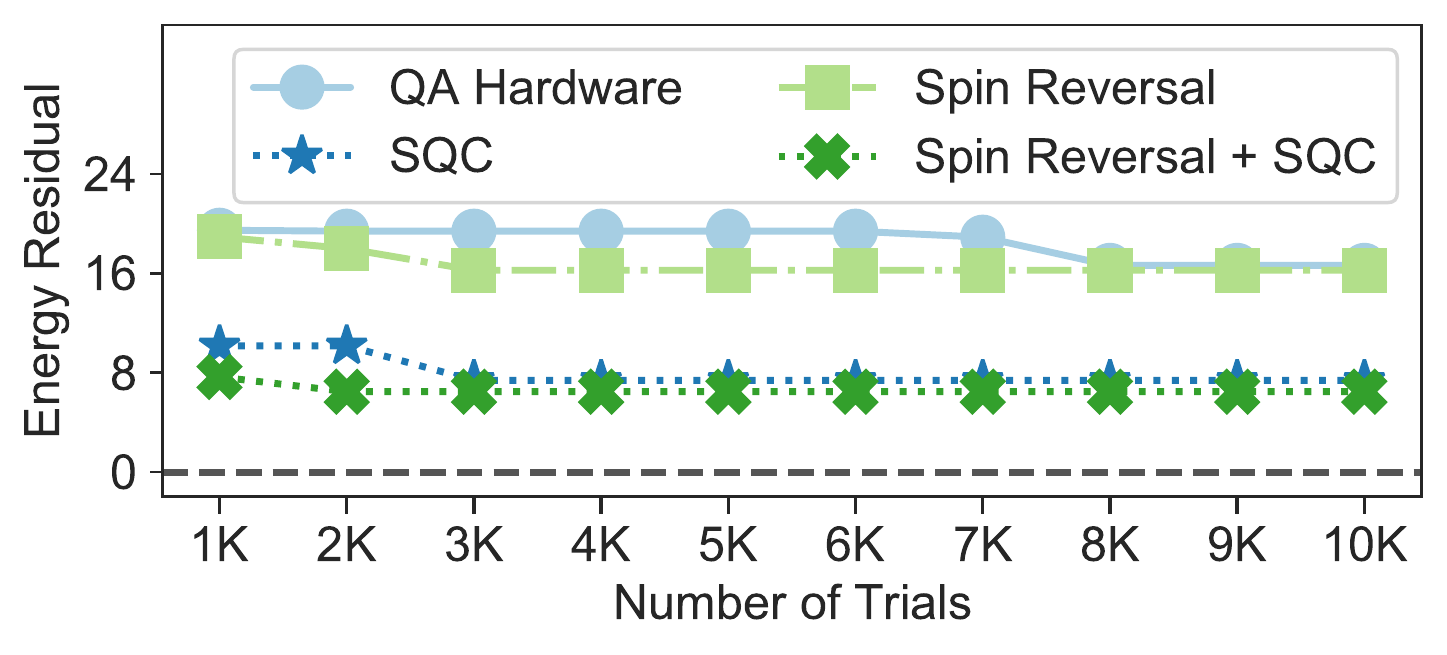}
	\caption{Energy Residual of a random benchmark with Spin Reversal Transform and Single Qubit Correction.  
	}
	\vspace{-0.05 in}
	\label{fig:impactofisd}
\end{figure}  

\subsection{Inter-Sample Delay vs. Single Qubit Correction}
Increasing the Inter-Sample-Delay (ISD) which increases the delay between successive QA reads to reduce the inter-sample correlations. 
Figure~\ref{fig:impactofisd} shows the ER of a benchmark executed on D-Wave QA under default ISD and when the ISD is increased. We also study the performance in the presence of SQC. We observe that similar to spin-reversal transform, while longer ISD reduces the ER, SQC outperforms and can be combined for further benefits.

\begin{figure}[htp]
	\centering
	\includegraphics[width=\columnwidth]{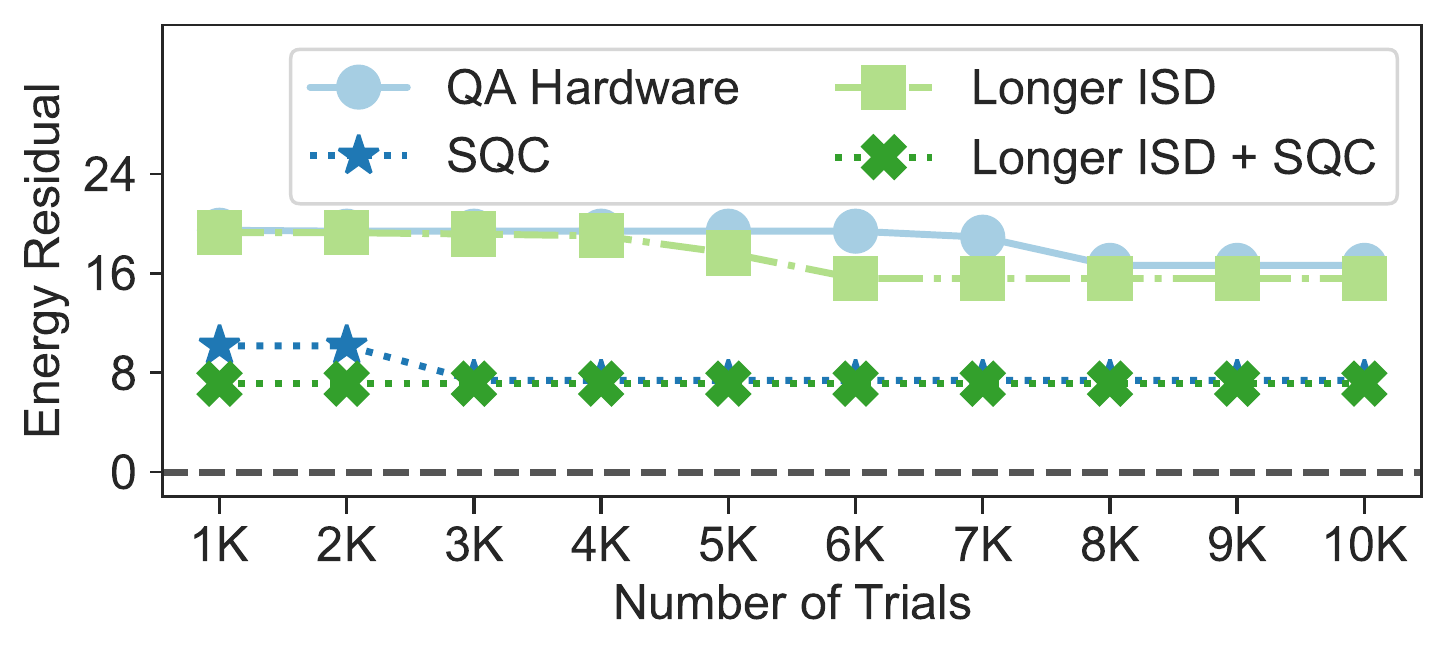}
	\caption{Energy Residual of a random benchmark with longer ISD and Single Qubit Correction.  
	\vspace{-0.1 in}
	}
	\label{fig:impactofisd}
\end{figure}  

We leverage the insights from this study to investigate if EQUAL can benefit from post-processing too. The additional advantage of relying on post-processing schemes is that it does not add any overheads in the casting or embedding step.

\section*{Acknowledgements}

Ramin Ayanzadeh was supported by the NSF Computing Innovation Fellows (CI-Fellows) program. 
Poulami Das was supported by the Microsoft Research PhD fellowship. 
This research was partially supported by the Office of the Vice Chancellor for Research and Graduate Education at the University of Wisconsin–Madison with funding from the Wisconsin Alumni Research Foundation.

%

\bibliographystyle{plain}
\bibliography{references}

\end{document}